\documentclass[aps,notitlepage]{revtex4-1}
\usepackage{amsfonts,amssymb,amsmath,bm}
\usepackage{graphicx,epsfig,color}
\usepackage{amssymb,amsfonts,amsmath}
\newcommand{\bs}{\mathbf {s}}
\newcommand{\br}{\mathbf {r}}
\newcommand{\bx}{\mathbf {x}}

\newcommand{\bv}{\mathbf {v}}
\newcommand{\bn}{\mathbf {n}}

\begin{document}

\title{Reconnection dynamics and mutual friction in quantum turbulence}

\author{Jason Laurie$^1$, Andrew W. Baggaley$^2$}

\affiliation{$^1$Department of Physics of Complex Systems, Weizmann Institute of Science, 234 Herzl Street, Rehovot 76100, Israel\\ 
$^2$School of Mathematics and Statistics, University of Glasgow, 15 University Gardens,
Glasgow, G12 8QW, UK\\
}

\date{\today}

\begin{abstract}
We investigate the behaviour of the mutual friction force in finite temperature quantum turbulence in $^4$He, paying particular attention to the role of quantized vortex reconnections. Through the use of the vortex filament model, we produce three experimentally relevant types of vortex tangles in steady-state conditions, and examine through statistical analysis, how local properties of the tangle influence the mutual friction force. Finally, by monitoring reconnection events, we present evidence to indicate that vortex reconnections are the dominant mechanism for producing areas of high curvature and velocity leading to regions of high mutual friction, particularly for homogeneous and isotropic vortex tangles. 
\end{abstract}
\keywords{quantum turbulence, mutual friction, quantum vortices, vortex reconnections}

\maketitle

\section{Introduction}
At temperatures above the lambda transition point $T_{\lambda}\approx2.17~\mathrm{K}$, liquid $^4$He behaves as a classical (normal) fluid with its dynamics described by the Navier-Stokes equations. However, at temperatures below $T_\lambda$, $^4$He becomes superfluid (known as helium II), and consequently, the classical Navier-Stokes description breaks down. 
A phenomenological model, which describes the motion of superfluid helium is the the two-fluid model of Landau and Tisza~\cite{tisza_theory_1947,landau_theory_1941,landau_theory_1949}. Here helium II is described as a mixture of two components, namely the superfluid and normal fluid. Each fluid is associated with separate velocity and density fields, denoted $\bv_n$ and $\rho_n$ for the normal fluid and
$\bv_s$ and $\rho_s$ for the superfluid respectively; the total density of helium II is $\rho=\rho_n+\rho_s$. These two components interact via mutual friction arising through the presence of extremely thin quasi-one-dimensional topological defects known as quantized vortices which can be excited in the superfluid component.  Unlike classical vortices, quantized vortices have a fixed vortex core size of $a\simeq 1~\AA$ (for superfluid $^4$He) and invoke an inviscid superfluid flow around them of fixed circulation defined in integer multiples of $\Gamma = h/m$, where $h$ is Planck's constant and $m$ is the mass of a $^4$He atom. For temperatures below $T_\lambda$, the relative ratio of the normal and superfluid components is temperature dependent, going from a pure normal fluid at $T=T_\lambda$ to a pure superfluid in the zero temperature limit.

Quantum or superfluid turbulence is the name given to the study of the chaotic (turbulent) fluid motions, which can be readily generated in quantum fluids such as superfluid $^4$He~\cite{vinen_quantum_2002,vinen_introduction_2006}. As first envisaged by Feynman, this manifests itself as a tangle of quantized vortices. Classical turbulent flows \cite{frisch_turbulence:_1995} are associated with the formation of eddies (vortices) that can appear at multiple scales. The quantum analogue is more interesting due to the fact that vorticity only emerges through quantized vortices of fixed size and circulation. For instance, large-scale topological features of the flow can be created via the formation of polarized vortex bundles, mimicking the large-scale classical eddies~\cite{baggaley_importance_2012,baggaley_vortex-density_2012}.  It is thus a hope, that the discreteness of vorticity in quantum turbulence may provide some insight into classical turbulence. Whether this proves to be true or not is still open, nonetheless quantum turbulence provides a rich field of study in its own right.

A major challenge in quantum turbulence is in the visualization of both the flow and the quantized vortex tangle, due to the extremely low temperatures involved. Hence, numerical methods become necessary to aid our understanding, providing a guide for both experiments and theory.
At present, many numerical quantum turbulence studies have focused on the behaviour of the superfluid component~\cite{adachi_steady-state_2010}, with direct comparisons to classical fluid turbulence~\cite{baggaley_vortex-density_2012}.  Several works have gone further and considered how quantum turbulence, at finite temperature, may effect the behaviour of the normal fluid component. However it is evident that the complexity associated with the two-way coupling of both the normal and superfluid components has limited theoretical progress. For example in some articles the two-fluid system is modelled using the Hall-Vinen-Bekarevich-Khalatnikov (HVBK) equations~\cite{hall_rotation_1956,bekarevich_phenomenological_1961}. However, the HVBK equations consider a coarse-grained superfluid vorticity in order to have a self-consistent two-fluid description at scales larger than the mean inter-vortex spacing, meaning that small-scale dynamics such are quantized vortex reconnections are ignored~\cite{henderson_nonlinear_1995,roche_quantum_2009}. Such reconnections events are discrete, dramatic events that are thought to be key to understanding quantum turbulence~\cite{bewley_characterization_2008,skrbek_developed_2012}. Therefore, vortex filament methods may be the more appropriate model. Unfortunately, the application of the vortex filament model coupled to the Navier-Stokes equations is a computationally expensive numerical scheme that has been currently limited to relatively simple configurations such as a single propagating vortex ring~\cite{kivotides_triple_2000} or for a limited number of reconnections between vortex rings~\cite{kivotides_superfluid_2001}.

Initial work by H\"{a}nninen~\cite{hanninen_dissipation_2013} has showed that a vortex reconnection leads to an enhancement of energy transfer through mutual friction.  The aim of this article is to try to examine the structure of the mutual friction force and to understand the role of vortex reconnections in a fully turbulent quantum tangle.  To achieve this, we use the vortex filament model of Schwarz~\cite{schwarz_three-dimensional_1985} to simulate the superfluid component only; for simplicity we only consider an imposed normal fluid velocity field, neglecting the back-reaction of the superfluid onto the normal fluid. We perform a series of numerical simulations, each with a different externally imposed normal fluid velocity profile. This produces a set of quantized vortex tangles, each with their own specific characteristics, which we shall use to investigate the behaviour of the corresponding mutual friction force. 

The outline of this articles is as follows: In Section~\ref{sec:vfm}, we introduce the vortex filament method used to model finite temperature superfluid $^4$He, and define the the formula for the mutual friction force which couples the superfluid component to the normal fluid. In Section~\ref{sec:setup} we describe the three numerical simulations we perform, giving precise details on how each quantum turbulence tangle is generated. We discuss our results in Section~\ref{sec:results}, before finally concluding in Section~\ref{sec:conclusions}.

\section{The vortex filament model}
\label{sec:vfm}

To simulate finite temperature quantum turbulence we utilize the well-known vortex filament method of Schwarz~\cite{schwarz_three-dimensional_1985}, where the description of the superfluid component entails modelling the motion of quantized vortex lines as one-dimensional space curves, $\bs(\xi,t)$, which evolve according to
\begin{subequations}\label{eq:vfm}
\begin{equation}\label{eq:Schwarz}
\frac{{\rm d}{\bf s}}{{\rm d}t}=\bv_s+\alpha \bs' \times (\bv_n-\bv_s)
-\alpha' \bs' \times \left[\bs' \times \left(\bv_n-\bv_s\right)\right],
\end{equation}
where $t$ is time, $\bs'={\rm d}\bs/{\rm d}\xi$ is the unit
tangent vector at the point $\bs$, $\xi$ is arc length, and
$\bv_n$ is the normal fluid velocity at the point $\bf s$.  Mutual friction effects with the normal fluid component are included through the last two terms in Eq.~(\ref{eq:Schwarz}) where $\alpha$ and $\alpha'$ are the non-dimensional temperature dependent friction coefficients for the vortex filament model.

The velocity of the superfluid component $\bv_s$ can be decomposed in terms of a self-induced velocity generated by the tangle $\bv_s^{\rm si}$, and an external superfluid flow $\bv_s^{\rm ext}$ such that 
$\bv_s=\bv_s^{\rm si}+\bv_s^{\rm ext}$. 
Here, the self-induced velocity $\bv_s^{\rm si}$ of the vortex line at the point $\bs$, is computed using the Biot-Savart law~\cite{saffman_vortex_1992}
\begin{equation}\label{eq:BS}
\bv_s^{\rm si} (\bs,\,t)=
\frac{\Gamma}{4 \pi} \oint_{\cal L} \frac{(\br-\bs) }
{\vert \br - \bs \vert^3}
\times {\rm {\bf d}}\br,
\end{equation}
\end{subequations}
where $\Gamma=9.97 \times 10^{-4}~\rm cm^2/s$
(in $^4$He) and the line integral
extends over the entire vortex configuration $\mathcal{L}$.

In the fully coupled system, the normal fluid velocity $\bv_n$ is determined by the three-dimensional Navier-Stokes equations coupled to the superfluid component through the mutual friction force ${\bf F}_{ns}$ (Eq.~(\ref{eq:mutualfriction})) that is included on the right-hand side of the Navier-Stokes momentum equation.  However, to obtain a vortex line density comparable to experimental work using the fully coupled model is currently beyond available computational resources. Nevertheless, we are still motivated to try and understand how a (turbulent) tangle of quantized vortices may perturb the normal fluid velocity field.

The mutual friction force ${\bf F}_{ns}$ is only created at the cores of the quantized vortex lines, which for the vortex filament model is expressed as a line integral taken along the quantum vortex tangle:
\begin{equation}\label{eq:mutualfriction}
{\bf F}_{ns}(\bx) =  \oint_{\mathcal{L}} \left\{\left(\rho_n\Gamma-D'\right)\left[{\bs}' \times \left( \bv_n - \dot{\bs}\right)\right] + D{\bs}' \times \left[{\bs'}\times \left( \bv_n - \dot{\bs}\right)\right]\right\}\delta(\bx - \bs)\ {\rm d}\xi({\bf s}).
\end{equation}
Here $D$ and $D'$ are the temperature dependent mutual friction coefficients that are experimentally measured~\cite{donnelly_observed_1998} and can be related to $\alpha$ and $\alpha'$~\cite{idowu_equation_2000}.

\section{The numerical setup}
\label{sec:setup}

All our simulations are performed at a single experimentally feasible temperature of $T=1.9~\rm K$, where the normal and superfluid densities are $\rho_n=6.10\times 10^{-2}~\rm g/cm^3$ and $\rho_s=8.44\times 10^{-2}~\rm g/cm^3$, leading to a ratio of $\rho_n/\rho_s\simeq 0.723$. At this temperature the mutual friction coefficients used in Eqs.~(\ref{eq:Schwarz}) and~(\ref{eq:mutualfriction}) are $\alpha=2.1\times 10^{-1}$, $\alpha'=8.3\times 10^{-3}$, $D=4.71\times 10^{-5}~\rm g/cm\, s$ and $D'=1.35\times 10^{-4}~\rm g/cm\, s$ respectively.

The numerical technique of the vortex filament model is to discretize the vortex lines by a series of points $\bs_j$ where $j=1,\dots,N$ held at a minimum distance of $\Delta\xi = 8\times 10^{-4}~\rm cm$. The Schwarz equation~(\ref{eq:Schwarz}) is evolved using a fourth-order Adams-Bashforth time-stepping method, with a typical timestep of $\Delta t= 5\times 10^{-5}~\rm s$ inside a periodic domain of size $\mathcal{D}=0.1~{\rm cm} \times 0.1~{\rm cm} \times 0.1~{\rm cm}$. Details on how the Biot-Savart integral~(\ref{eq:BS}) is regularized, how a tree-approximation is utilized (with an opening angle of $0.3$) to improve computational efficiency, and how vortex reconnections are artificially implemented are all described in~\cite{baggaley_vortex-density_2011,baggaley_tree_2012}.

We perform three numerical simulations, each with a different forcing mechanism to induce a particular structure of quantum vortex tangle. The simulations are initialised with $6$ randomly positioned vortex rings to seed a vortex tangle. We have chosen the forcing parameters in each simulation specifically to ensure that the vortex line density $L=\Lambda/V$, where $\Lambda = \int_{\mathcal{L}}\, d\xi({\bf s})$ is the total vortex line length and $V= 1\times 10^{-3}~{\rm cm}^3$ is the volume of the domain, saturates at approximately the same value.  This gives us a standard reference to make direct comparisons between the three vortex tangles.  Below, we outline our three numerical setups.

\subsection{Generation of a Vinen tangle}

For the first simulation we periodically inject randomly orientated vortex rings of radius $R=1.91\times 10^{-2}~\rm cm$ directly into the superfluid component.  We choose an injection period of $6\times 10^{-4}~\rm s$ to ensure that a dense vortex tangle is produced.  The simulation is evolved so that a non-equilibrium statistical steady state regime is achieved through the balance of the periodic injection of vortex rings and dissipation by mutual friction. This is confirmed through the observation of statistical stationarity of the vortex line density which saturates at $L\simeq 2.0\times 10^{4}~\rm cm^{-2}$. In this setup, the normal fluid velocity profile is considered to be at rest ($\bv_n=0$). 

\subsection{Generation of a Counterflow tangle}

The second simulation is a counterflow setup, where the quantum vortex tangle is excited through a uni-directional normal fluid flow $\bv_n = U{\bf e}_x$, with speed $U=1~\rm cm/s$.  In counterflow experiments, the mean normal fluid flow invokes an oppositely orientated superfluid flow of $\bv_s^{\rm ext}=-\left(\rho_n/\rho_s\right)\bv_n$ in order to satisfy overall mass conservation of the fluid.  In this respect, we also impose this external superfluid flow. The simulation is performed so that the vortex line density saturates, reaching a value of $L\simeq 2.1\times 10^{4}~\rm cm^{-2}$.

\subsection{Generation of a Polarized tangle}

Finally, for the third simulation we use a normal fluid profile $\bv_n$ obtained from a numerical snapshot of classical homogeneous and isotropic turbulence generated by the Navier-Stokes equations taken from the John Hopkins Turbulence Database~\cite{li_public_2008}.  The dataset consists of a velocity field discretized on a mesh of $1024^3$ points. The estimated Reynolds number of the velocity snapshot is ${\rm Re}\sim \left(L_0/\eta_0\right)^{4/3}\simeq 3025$ where $L_0$ and $\eta_0$ are the integral and Kolmogorov scales respectively. The reason for using a single stationary snapshot for the normal fluid velocity profile and not a time-dependent one is based solely on computational constraints. As with the counterflow setup, the normal fluid flow excites a quantum vortex tangle which after a transition period reaches a non-equilibrium statistical stationary state monitored through the vortex line density $L\simeq 2.0\times 10^{4}~\rm cm^{-2}$.

\section{Results}
\label{sec:results}
\subsection{Tangle structure}

\begin{figure}[htbp]
	\centering
	\includegraphics[height=3.4cm]{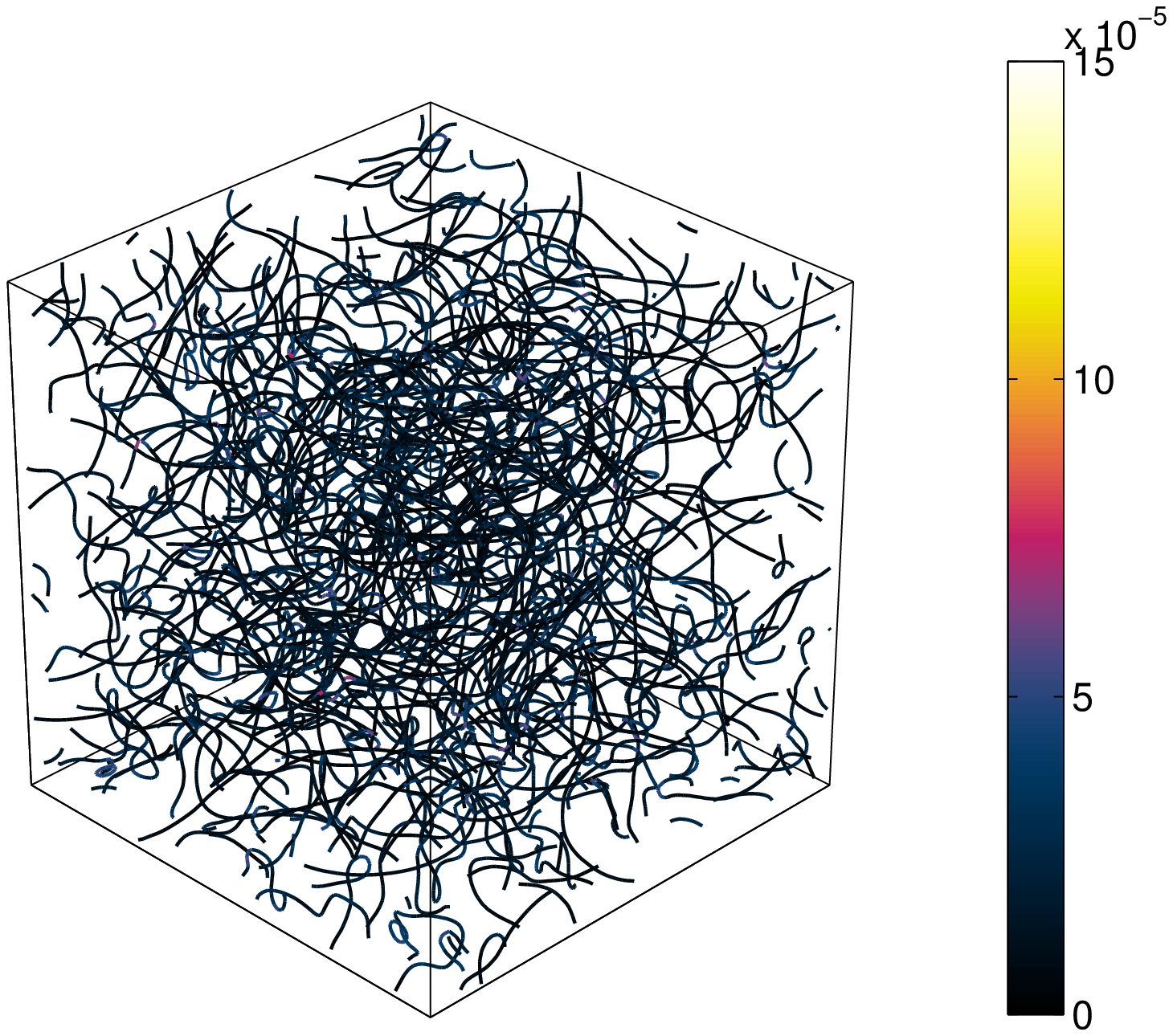}
	\hfill
	\includegraphics[height=3.4cm]{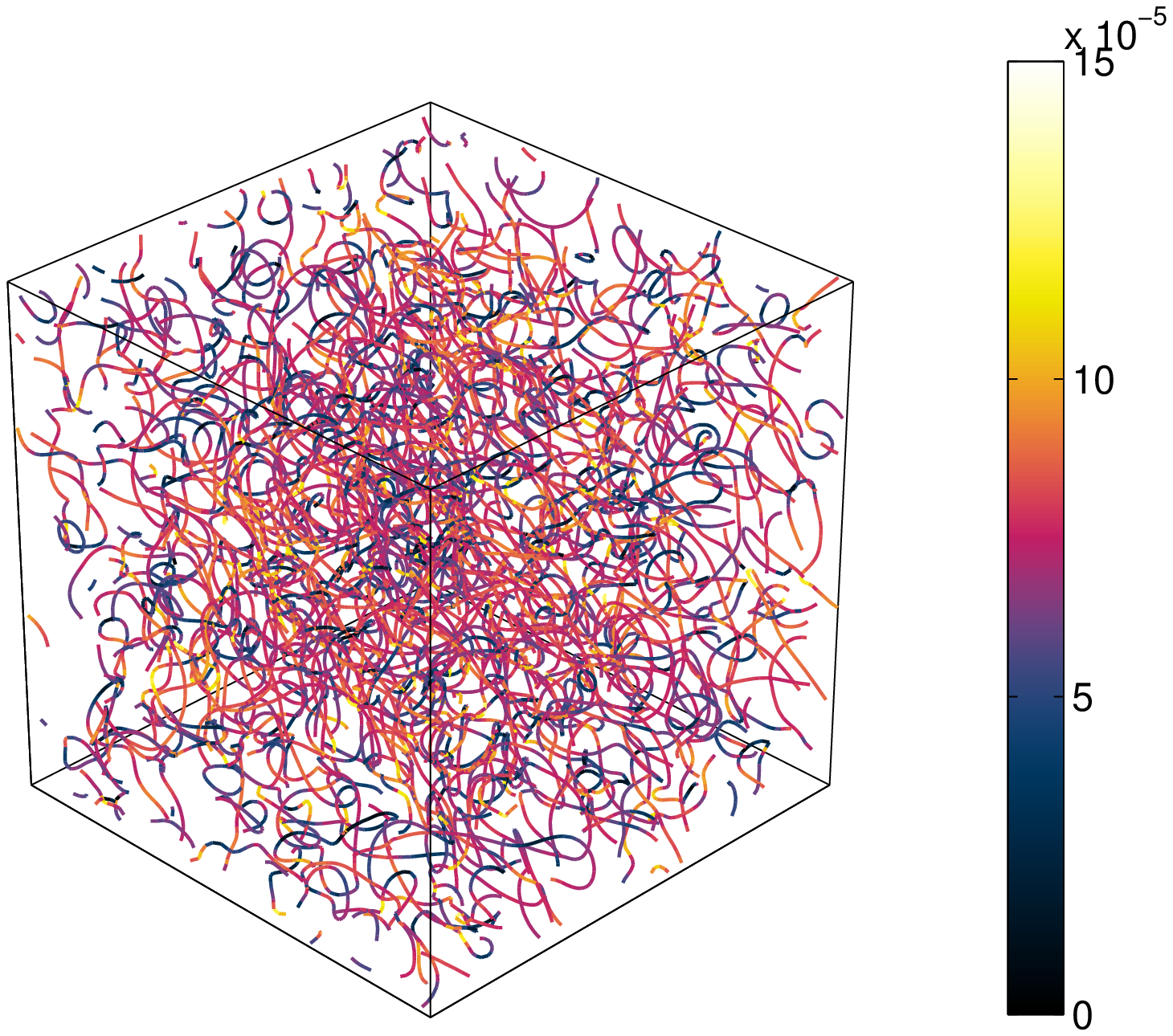}
	\hfill
	\includegraphics[height=3.4cm]{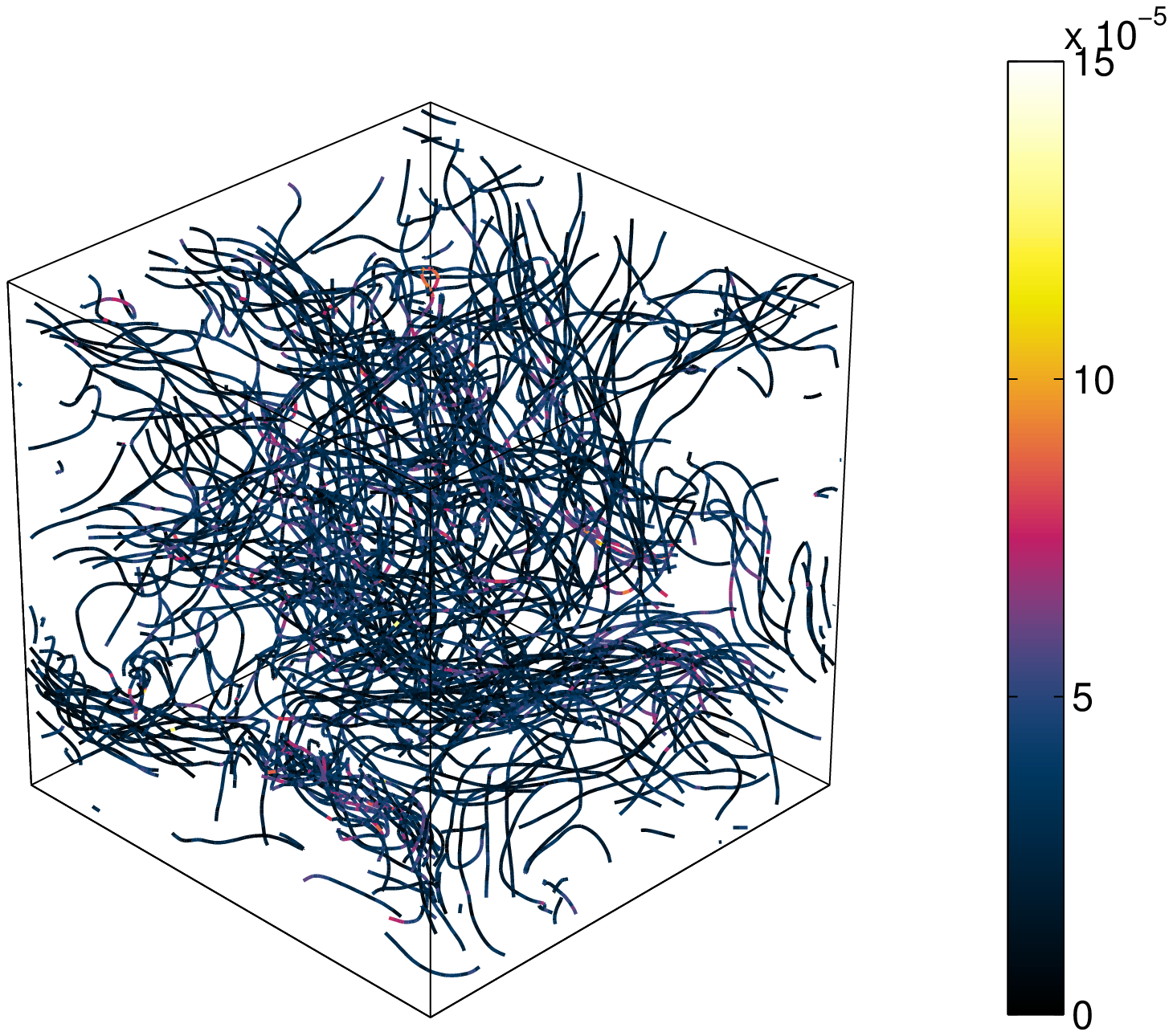}
	\caption{Vortex tangle snapshots, Vinen (left), Counterflow (middle) and Polarized (right) from the three numerical simulations.  The color code indicates the magnitude of the mutual friction $\left|{\bf F}_{ns}(\bx)\right|$ given by Eq.~(\ref{eq:mutualfriction}).}
	\label{fig:tangle}
\end{figure}

In Fig.~\ref{fig:tangle} we present images of all three vortex tangles taken during stationary conditions.  Each vortex line segment is colour-coded to its respective magnitude of the mutual friction force~(\ref{eq:mutualfriction}). The most evident distinction between the three is in the tangle structure itself.  Both the Vinen (left) and the Counterflow (centre) tangles show a distinct lack of large-scale features; the tangles appear homogeneous and completely random in orientation. On the other hand, the third tangle (Polarized, right) displays clear large-scale features in the form of polarized vortex bundles consisting of multiple quantized vortex lines orientated in the same direction. In fact, as was noticed by Kivotides~\cite{kivotides_coherent_2006}, who also used a stationary normal fluid profile, the polarized vortex bundles are located precisely in the regions where the classical vortices appear in the normal fluid flow. The structure of the tangle plays an essential role in the reconnection dynamics as can be seen from the reconnection rates of our simulations: $5.43 \times 10^{4}~{\rm s}^{-1}$ (Vinen), $5.03 \times 10^{4}~{\rm s}^{-1}$ (Counterflow) and $3.31 \times 10^{4}~{\rm s}^{-1}$ (Polarized).  We find that reconnections are suppressed in the Polarized tangle by almost a factor of two compared to the other two simulations. Indeed, L'vov {\it et al.} hypothesized~\cite{lvov_bottleneck_2007} that the formation of polarized vortex bundles will likely lead to a reduction of reconnections. Furthermore, we notice that in both the unpolarized Vinen and Counterflow tangles the reconnection rates are almost identical, even though the forcing mechanisms are different in both cases. Therefore, we may well surmise that the structure of the tangle (random or polarized) strongly influences the reconnection rate in steady state conditions, for comparable vortex line densities.

From the colour scale (the same for all three tangles) we observe that the Vinen and Polarized tangles produce a much lower mean mutual friction when compared to the Counterflow tangle. However, on closer inspection of all three snapshots, one can find isolated vortex segments where the mutual friction is extremely high in comparison to the typical value. We shall see that these extreme values can be attributed to vortex reconnections.

\subsection{Bi-variate PDFs and vortex reconnections}

To probe deeper into the local structure of the vortex tangles we compute kernel density estimates~\cite{silverman_density_1986} for the bi-variate probability density distributions (PDFs) to check for correlations between local vortex line quantities such as the local mutual friction $\left| {\bf F}_{ns}\right|$, local curvature $\kappa=|\bs''|$, superfluid speed $\left|\dot{\bf s}\right|$, and the local vorticity production rate $P=-\kappa \dot{\bs}\cdot \bn$, where $\bn$ is the unit normal vector to the vortex segment. Note our numerical resolution $8 \times 10^{-4} ~ {\rm cm}<\Delta \xi < 1.6 \times 10^{-3}~ {\rm cm}$ implies the maximum curvature that can be resolved is $\kappa_{\rm max} \simeq 1250 ~ {\rm cm}^{-1}$.

We collect the data by averaging $20$ snapshots taken $2.5\times 10^{-3}~\rm s$ apart once in steady state conditions. The PDFs are shown using a logrithmic grey scale with black indicating regions of highest probability. On top of the PDFs, we display specific vortex points that have recently undergone a vortex reconnection event.  Theses are located by recording points that have experienced a reconnection within a small time window of $\Delta t_{\rm rec}=1\times 10^{-3}~\rm s$ from the time each snapshot was taken. The reason for this particular choice of $\Delta t_{\rm rec}$ is that one expects after a reconnection event Kelvin waves are generated at a scale of approximately the inter-vortex spacing, dynamically changing the local structure of the neighbouring vortex segments.  Hence, $\Delta t_{\rm rec}$ is computed to be the timescale of one Kelvin wave period $\Delta t_{\rm rec}\sim 2\pi/\omega_{k_\ell}$ defined at the inter-vortex scale $k_\ell = 2\pi/\ell = 2\pi/L^{-1/2}$ with frequency $\omega_k = (\kappa k^2/ 4\pi)\left[\ln\left(1/ka\right)-\gamma -3/2\right]$.  This enables us to capture the initial dynamics of the vortex segment post reconnection.  In Table~\ref{tab:mean} we present an overview of the mean values of each local quantity considered for the whole tangle and also for the reconnection events only.

\begin{table}[h!]
\begin{center}
\caption{Mean values of the mutual friction $\left| {\bf F}_{ns} \right|$, curvature $\kappa$, superfluid speed $|\dot{\bs}|$ and vorticity production $P$. Mean values of reconnection events are given in brackets.\label{tab:mean}}
\begin{tabular}{cccc}
\hline
Tangle (Recon)&Vinen & Counterflow & Polarized\\
\hline
$\left| {\bf F}_{ns} \right|~\left[  \mathrm{g}/ \mathrm{cm}^2\, \mathrm{s}^2 \right]$ & $1.55$ ($3.33$) $\times 10^{-5}$ & $7.12$ ($7.28$) $\times 10^{-5}$ & $2.77$ ($4.63$) $\times 10^{-5}$ \\
$\kappa~\left[\mathrm{cm}^{-1} \right]$ & $159.76$ ($516.21$) & $203.47$ ($755.03$) & $113.78$ ($478.11$)\\
$|\dot{\bs}|~\left[ \mathrm{cm}/\mathrm{s} \right]$ & $0.19$ ($0.47$) & $0.28$ ($0.65$) & $0.21$ ($0.51$)\\
${P}~\left[ \mathrm{s}^{-1} \right]$ & $-6.86$ ($-32.26$) & $5.22$ ($-72.55$) & $1.11$ ($-7.90$) \\
\hline
\end{tabular}
\end{center}
\end{table}

\begin{figure}[htbp]
	\centering
	\includegraphics[width = 0.32\textwidth]{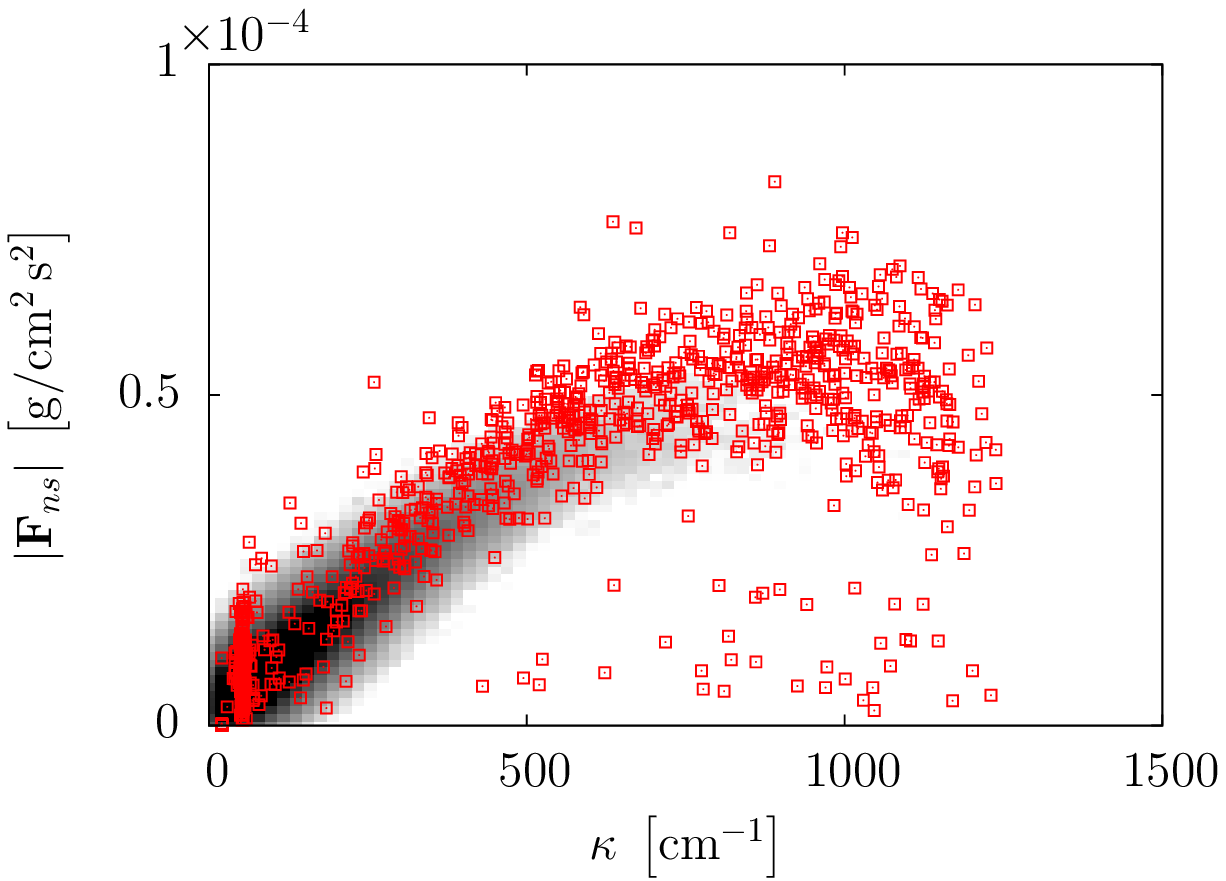}
	\includegraphics[width = 0.32\textwidth]{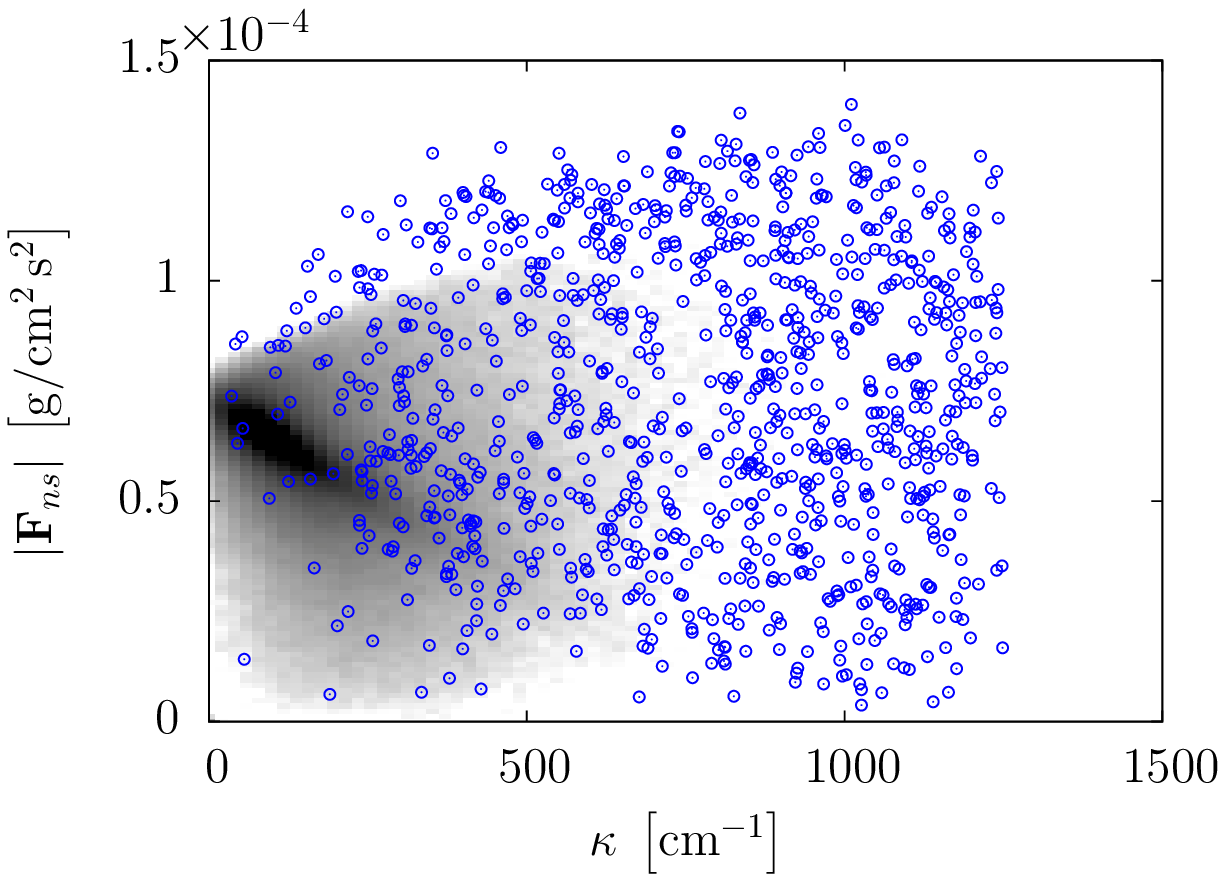}
	\includegraphics[width = 0.32\textwidth]{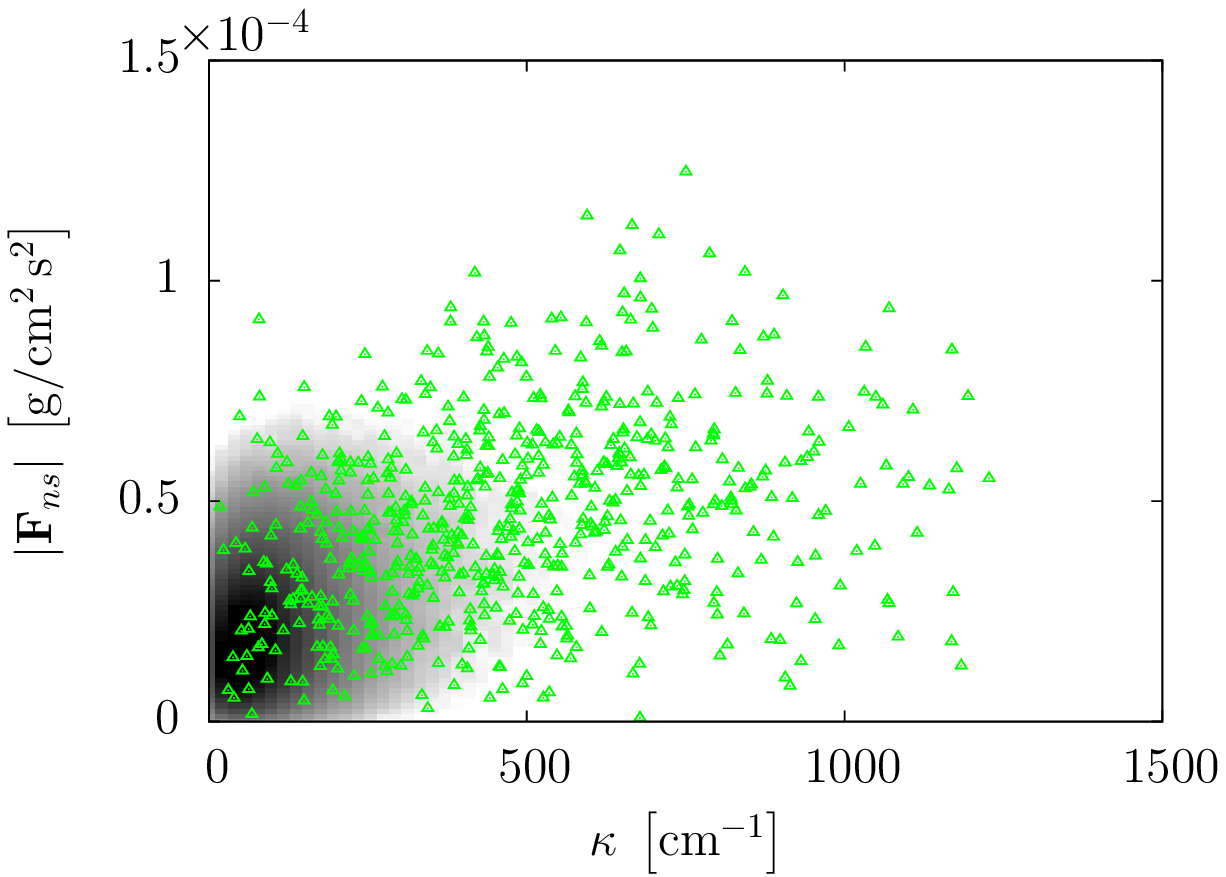}

	\caption{Kernel density estimates of the probability density function of the local mutual friction ${\bf F}_{ns}$ verses the local curvature $\kappa$ for the Vinen tangle (left), Counterflow tangle (centre) and Polarized tangle (right). The coloured symbols indicate reconnection events.}
	\label{fig:pdf_mf_curv}
\end{figure}

Figure ~\ref{fig:pdf_mf_curv} displays the PDF of the curvature verses the magnitude of the mutual friction force for each tangle. We observe that the typical vortex tangle segment has far lower curvatures than those that have recently undergone a reconnection.  This is also verified by the values given in Table~\ref{tab:mean} indicating that reconnections lead to approximately four times the mean curvatures observed in each tangle. Additionally, for the Vinen and Polarized tangles we see a positive correlation between the curvature of the point and the associated mutual friction force.  This indicates that for these tangles, the most important regions that may effect the normal fluid behaviour are those that have recently experienced a reconnection event. 

In contrast the presence of the non-zero mean normal fluid flow that quintessentially defines superfluid counterflow, means each vortex segment will produce a uniform drag against the normal fluid velocity. This is also indicated by the mean mutual friction force being far higher than compared to the Vinen and Polarized tangles, as seen in Fig.~\ref{fig:pdf_mf_curv} and Table~\ref{tab:mean}.  It is quite evident that vortex reconnection events lead to large local fluctuations in the mutual friction force. However as these fluctuations are on top of a large non-zero mean it is possible that vortex reconnections can reduce (as well as enhance) the local mutual friction.  

\begin{figure}[htbp]
	\centering
	
	\includegraphics[width = 0.32\textwidth]{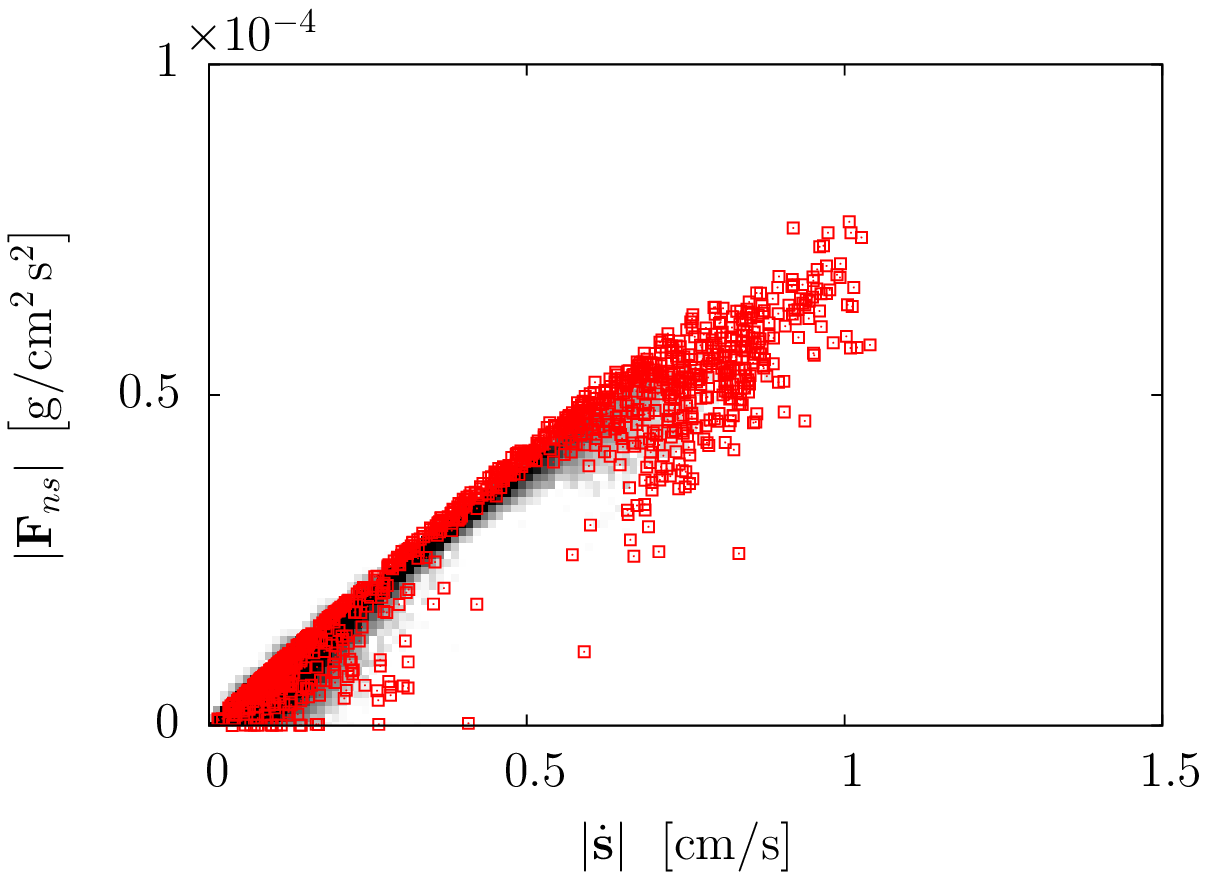}
	\includegraphics[width = 0.32\textwidth]{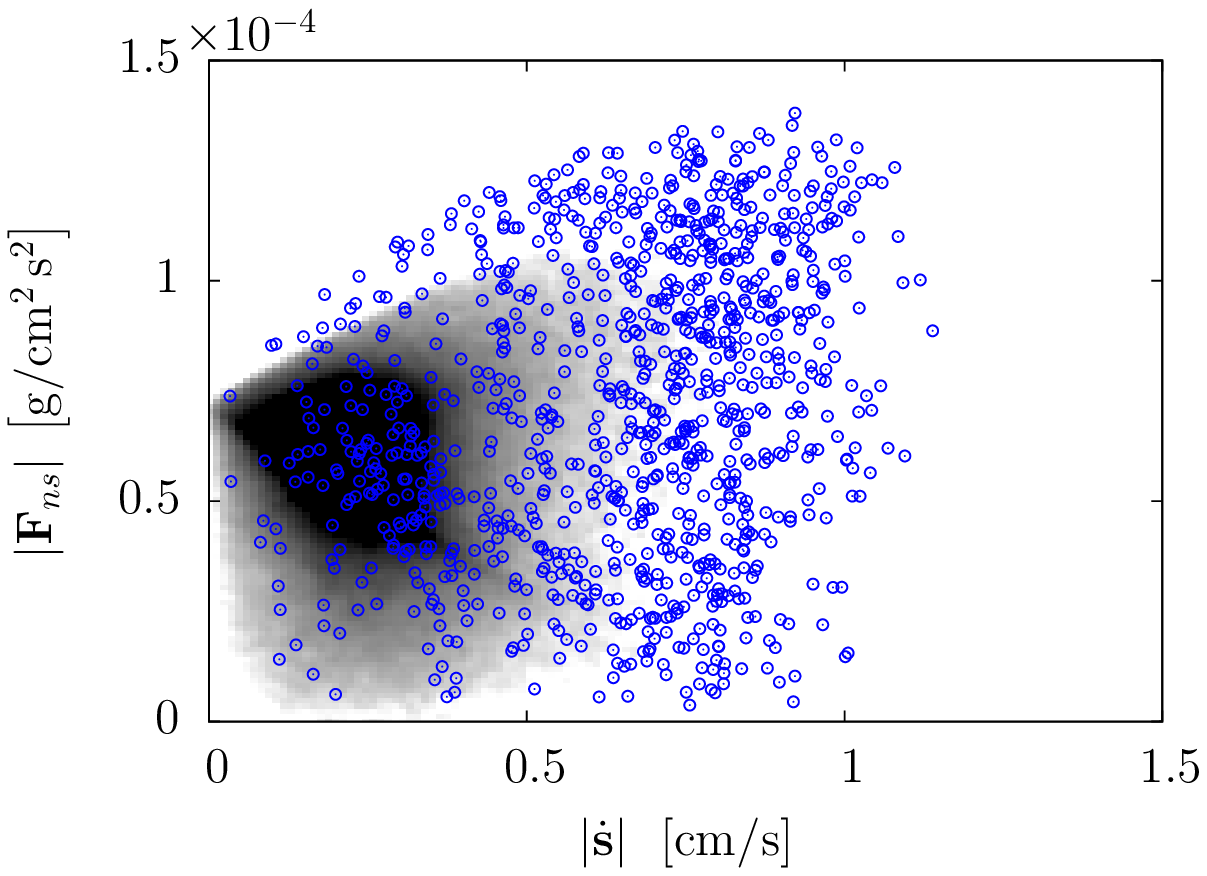}
	\includegraphics[width = 0.32\textwidth]{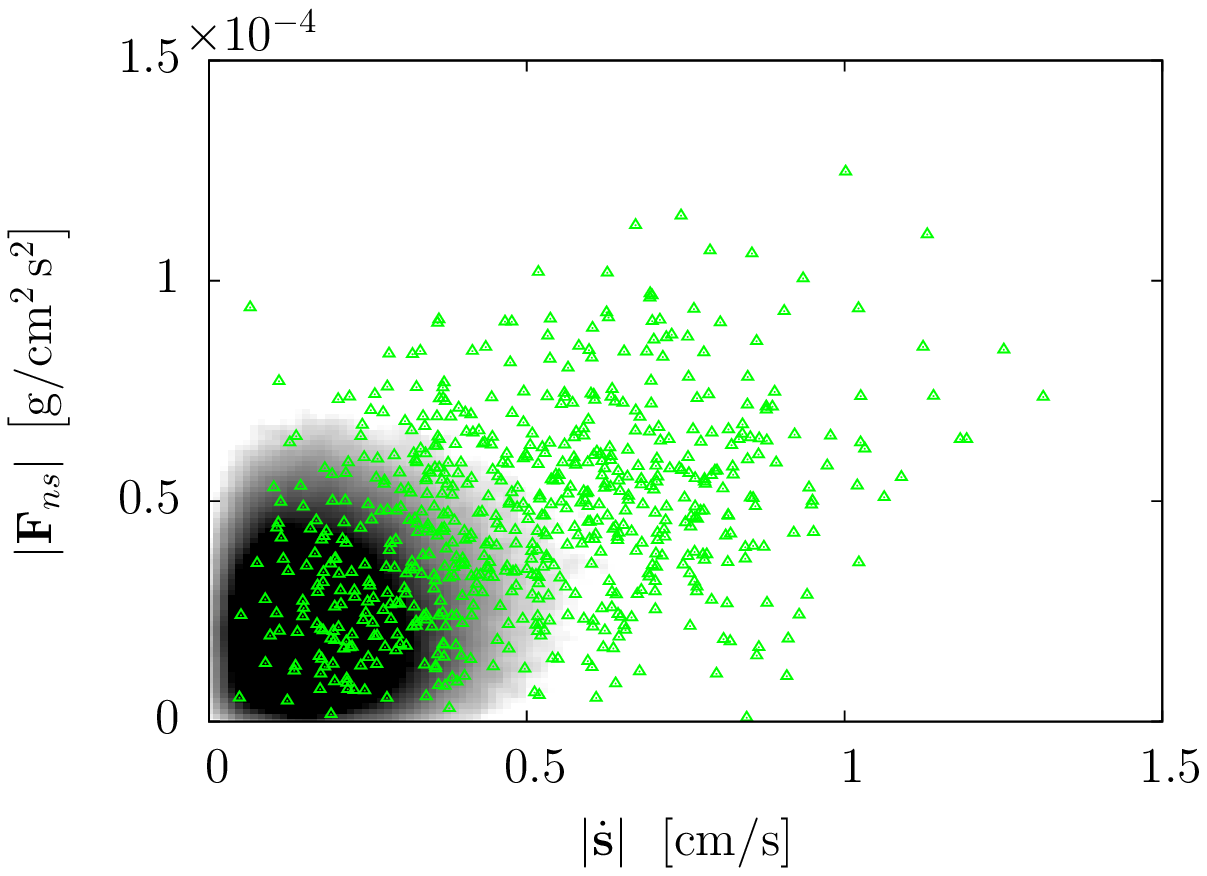}

	\caption{Kernel density estimates of the probability density function of the mutual friction ${\bf F}_{ns}$ verses the local superfluid speed $\left|\dot{\bf s}\right|$ for the Vinen tangle (left), Counterflow tangle (centre) and Polarized tangle (right). The coloured symbols indicate reconnection events.}
	\label{fig:pdf_mf_vel}
\end{figure}

From the definition of the mutual friction force~(\ref{eq:mutualfriction}), we see that its value is proportional to the velocity difference between the normal and superfluid components.  Due to the fact that the model Vinen tangle in our numerical simulations is coupled to a static normal fluid flow, we should expect that the mutual friction force will be correlated to the local speed of the vortex tangle. This is perfectly verified in Fig.~\ref{fig:pdf_mf_vel} by the stark positive correlation between the superfluid speed and mutual friction, especially indicated by the reconnection events. A similar behaviour, but with far more fluctuations, is also observed for the Polarized tangle.  These fluctuations will most likely be associated to the fluctuations of the turbulent normal fluid velocity.  We observe that the correlation is not as strong in the Counterflow tangle, presumably due to the larger mean mutual friction.

\begin{figure}[htbp]
	\centering
	\includegraphics[width = 0.32\textwidth]{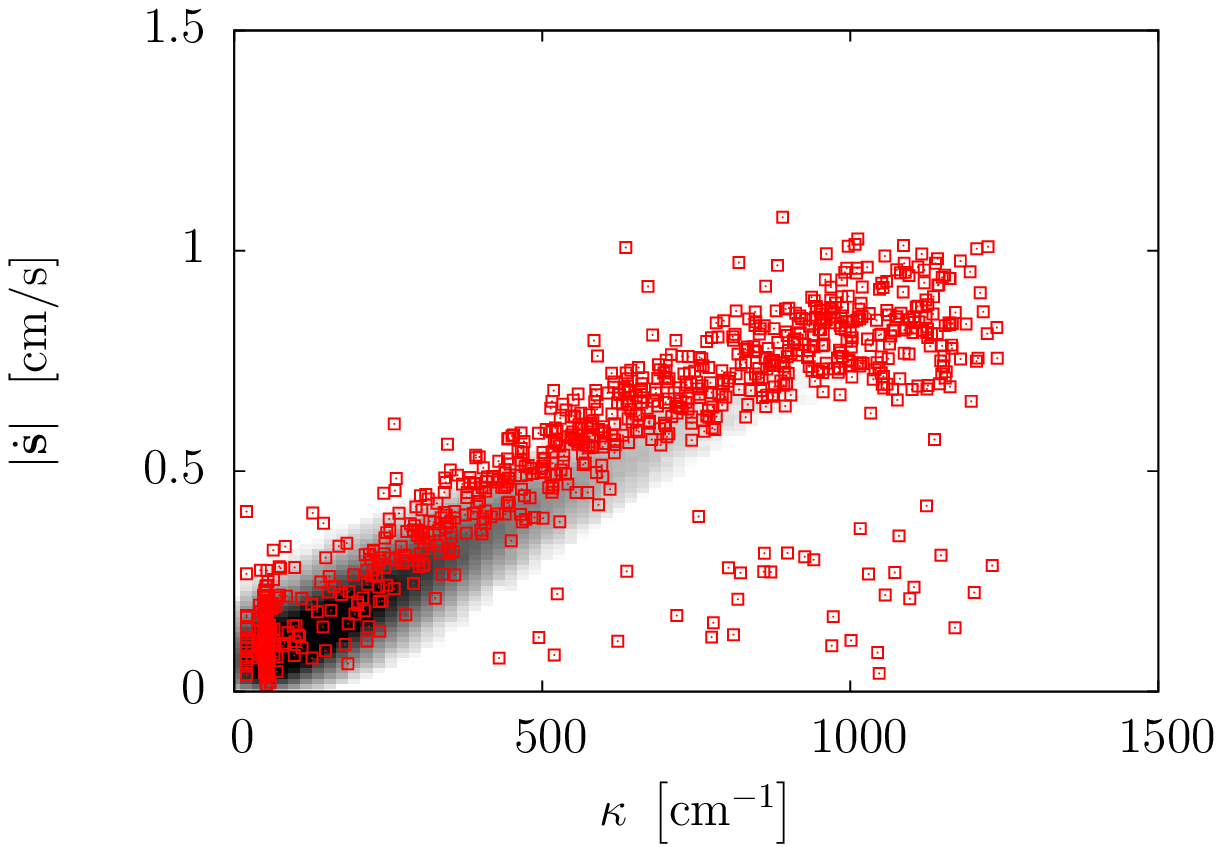}
	\includegraphics[width = 0.32\textwidth]{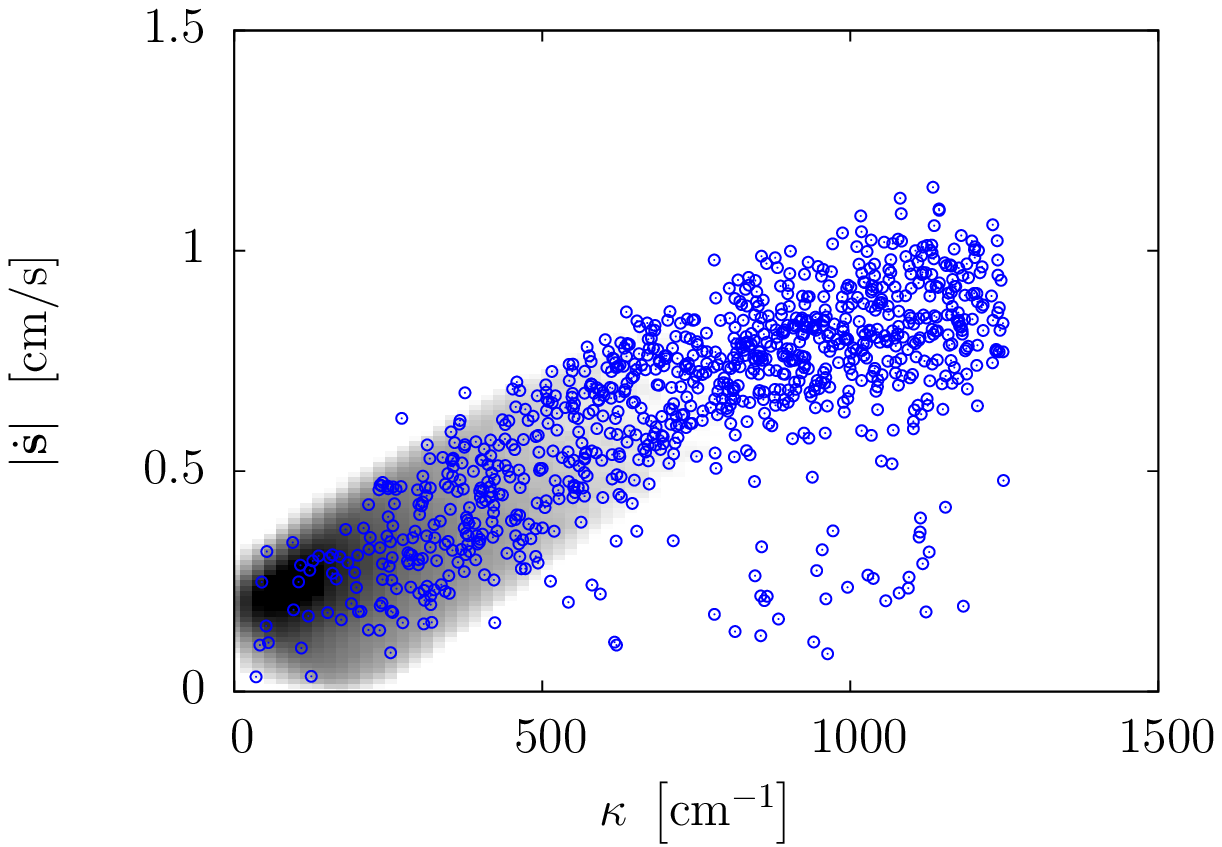}
	\includegraphics[width = 0.32\textwidth]{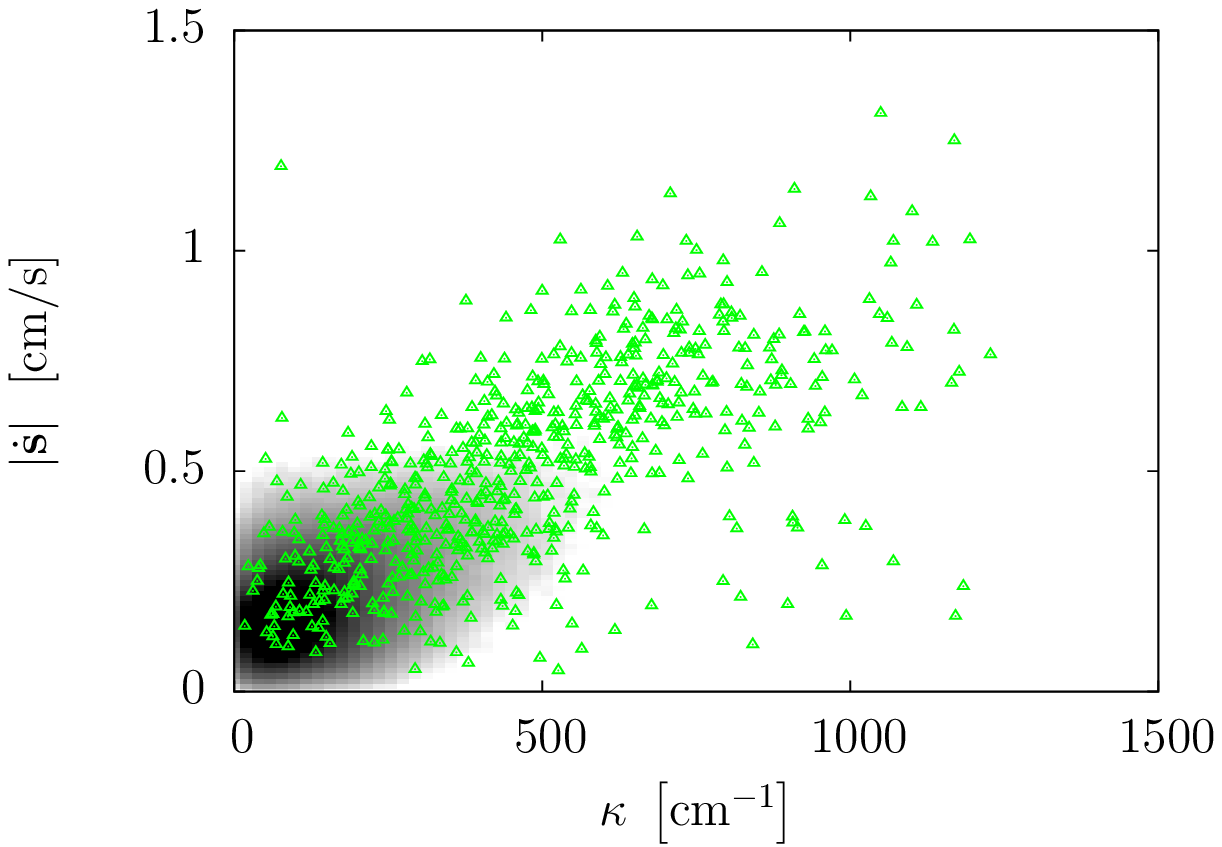}

	\caption{Kernel density estimates of the probability density function of the curvature $\kappa$ verses the local superfluid speed $\left|\dot{\bf s}\right|$ for the Vinen tangle (left), Counterflow tangle (centre) and Polarized tangle (right). The coloured symbols indicate reconnection events.}
	\label{fig:pdf_curv_vel}
\end{figure}

The almost identical distributions observed in Figs.~\ref{fig:pdf_mf_curv} and~\ref{fig:pdf_mf_vel} lead us to consider PDFs of curvature against superfluid speed.  These PDFs are presented in Fig.~\ref{fig:pdf_curv_vel} and demonstrate almost perfect correlation between the curvature and speed for the vortex reconnection events in all of the tangles. Indeed, we see that the majority of the vortex points within each tangle are of small curvatures and speeds with the rare and spontaneous reconnection events being responsible for the generation of high curvature and high superfluid velocities. Considering the data presented in Figs.~\ref{fig:pdf_mf_vel} and~\ref{fig:pdf_curv_vel} we can conclude that vortex reconnections produce areas of high curvature and high velocities which in turn will contribute to strong fluctuations of the mutual friction. It is important to note that the numerical resolution limits the maximum curvature we can resolve, and hence the maximum velocities produced after a reconnection.

If, like the Vinen and Polarized tangles, there is no mean normal fluid flow, reconnection events become the dominant mechanism for creating regions of high mutual friction. Indeed, one can imagine that the normal fluid will experience instantaneous intense `explosions' of mutual friction force in the neighbourhood of reconnections in the superfluid component. This type of force will also predominately act on small scales due to the localized nature of the vortex reconnection events. Therefore, one may question the potential for the mutual friction force to destabilise the normal fluid if it was to act sporadically at such small scales--we discuss this further in the conclusions.

\begin{figure}[htbp]
    \vspace{1cm}
	\centering
	\includegraphics[width = 0.25\textwidth]{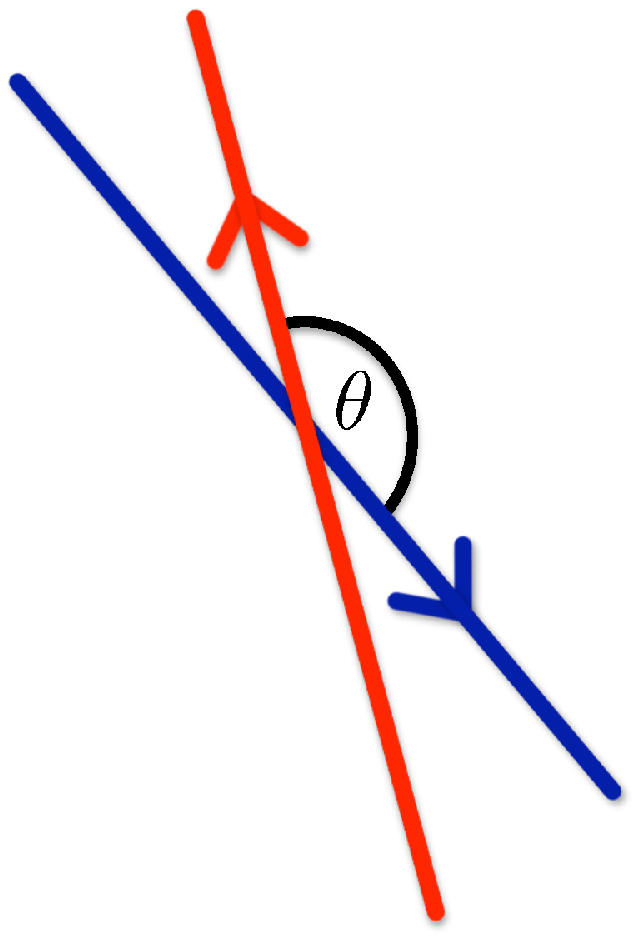}
	\includegraphics[width = 0.5\textwidth]{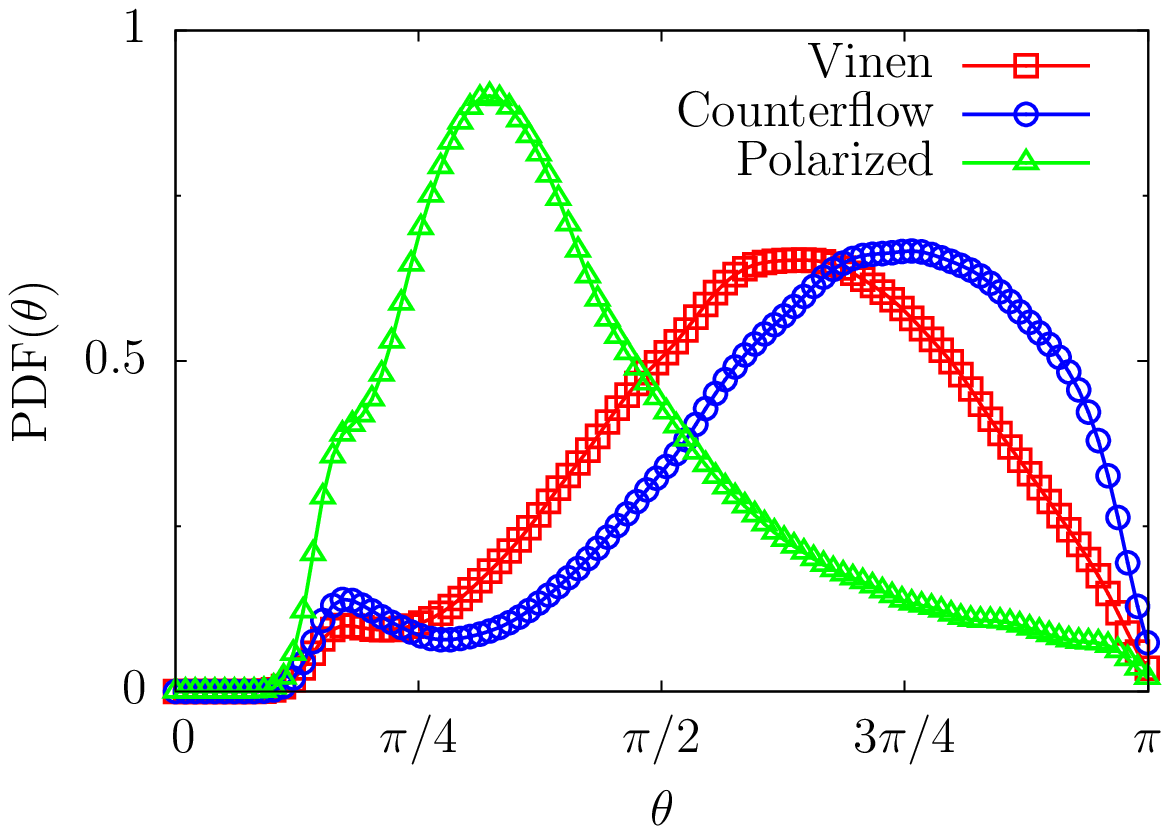}
	\caption{(left) Diagram showing the definition of the reconnection angle $\theta$ between two quantized vortex lines (red and blue).  The arrows point in the direction of the vorticity vector ${\bf s}'$. (right) Kernel density estimation of the probability density function of the reconnection angle $\theta$ of quantum vortex reconnections for the Vinen (red squares), Counterflow (blue circles) and Polarized (green triangles) vortex tangles. Here, $\theta=0$ corresponds to parallel vortex lines and $\theta=\pi$ to anti-parallel vortex lines.}
	\label{fig:angle}
\end{figure}

Let us comment on the particularly large curvatures observed in the Vinen and Counterflow tangles as opposed to those of the Polarized tangle, see Table~\ref{tab:mean}.  In Fig.~\ref{fig:angle}, we compute  PDFs of the reconnection angles of the reconnecting vortex segments for each tangle. We notice that both the Counterflow and Vinen tangles contain predominately large angle vortex reconnections corresponding to reconnections between almost anti-parallel vortex lines. From the schematic in Fig.~\ref{fig:angle} (left) one can easily understand how large angle ($\theta > \pi/2$) reconnections lead to regions of higher curvatures than those of small angles--for large angle reconnections two sharp (high curvature) cusps are created post reconnection. As previously discussed, there is a strong correlation between the local curvature and the local speed of the vortex segment. Therefore, large angle reconnections may lead to significantly higher velocities and higher mutual friction forces than those with smaller angles. From the data presented in Table~\ref{tab:mean}, one may argue that this is not necessarily the case for the Vinen tangle (as the values are similar to those of the Polarized tangle). However, one should be aware that there are many reconnections in the Vinen tangle occurring near $\kappa \simeq 50~\rm cm^{-1}$ as seen in Figs.~\ref{fig:pdf_mf_curv} and \ref{fig:pdf_curv_vel}.  This value of curvature corresponds to the that of the injected vortex rings $\kappa = 1/R \simeq 52.35~\rm cm^{-1}$ meaning we are recording instantaneous reconnections due to the random placement of vortex rings from our forcing mechanism.  Consequently, these small curvatures are biasing the actual average curvatures expected in a completely random Vinen tangle that one could produced through other means.

\begin{figure}[htbp]
	\centering
	\includegraphics[width = 0.32\textwidth]{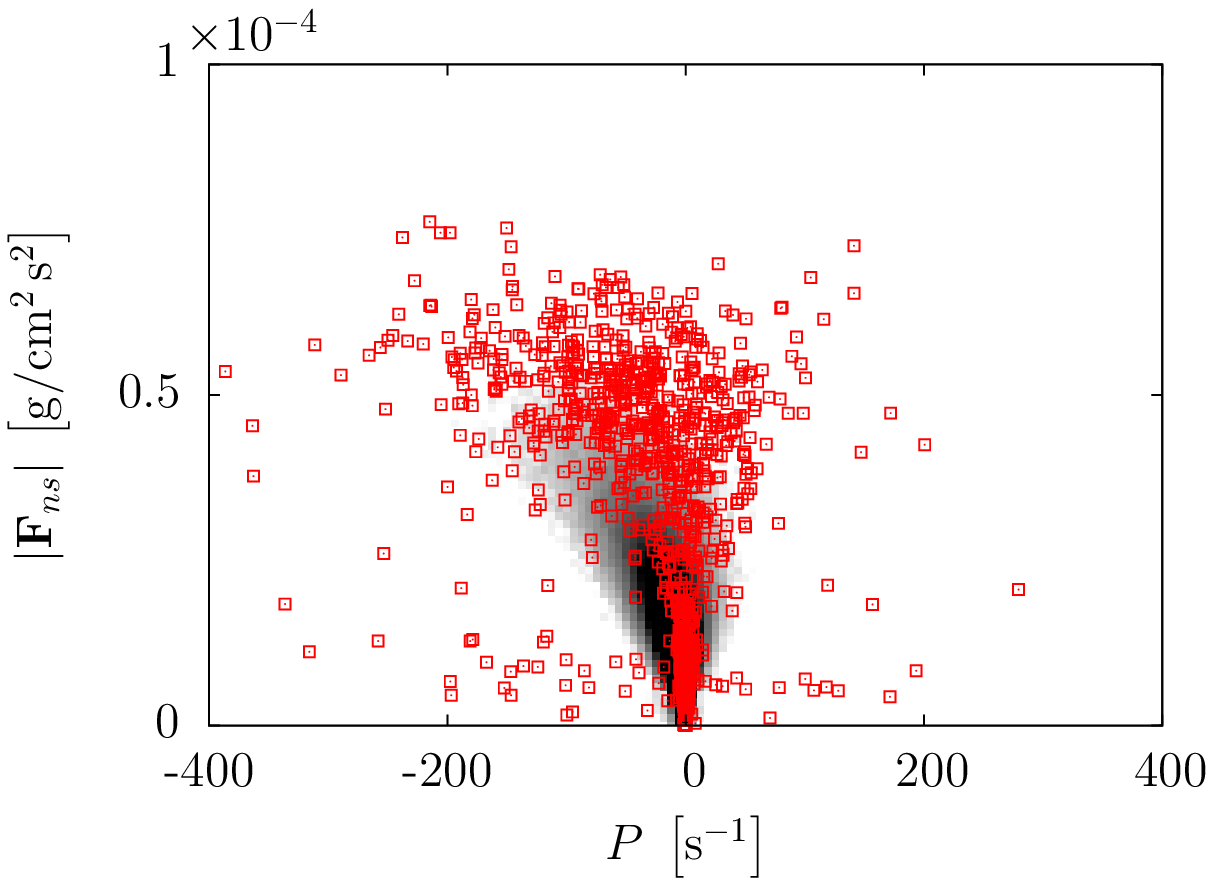}
	\includegraphics[width = 0.32\textwidth]{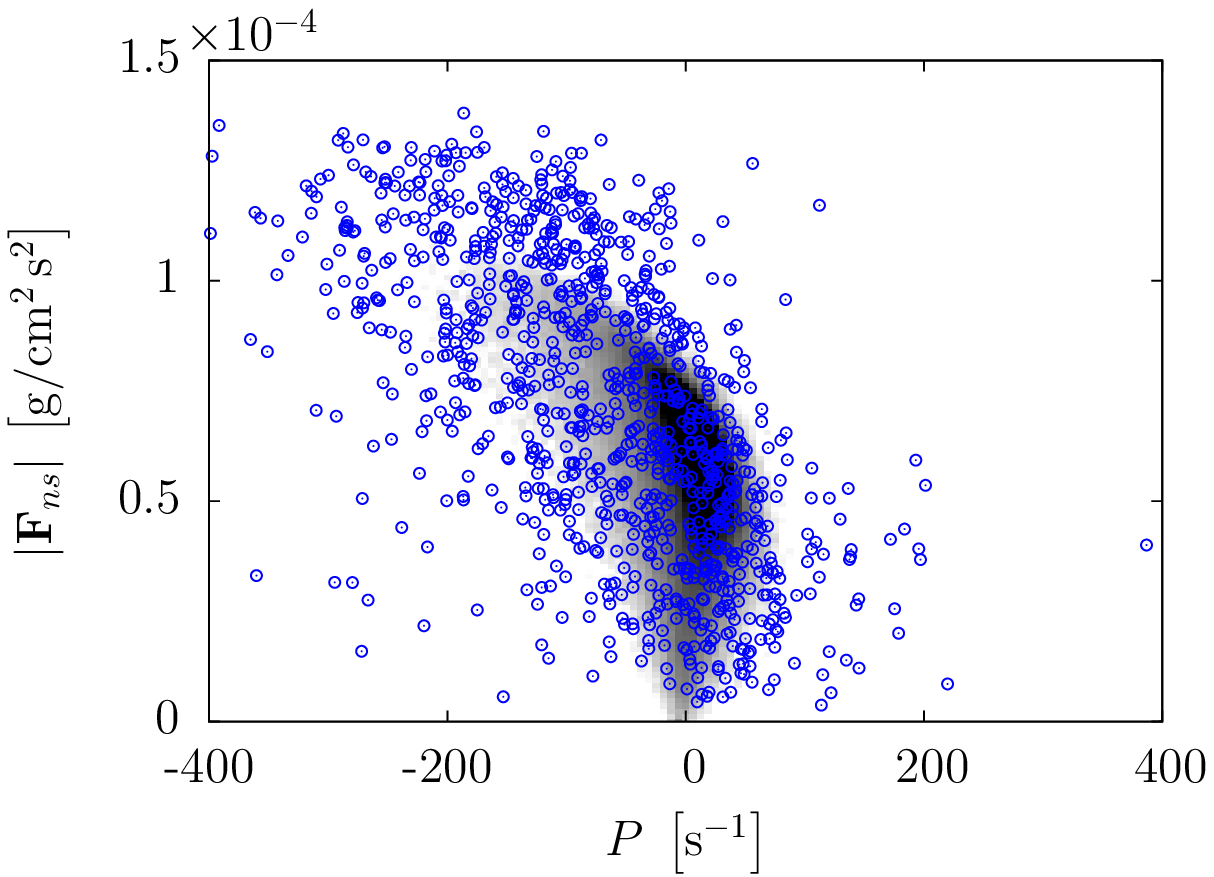}
	\includegraphics[width = 0.32\textwidth]{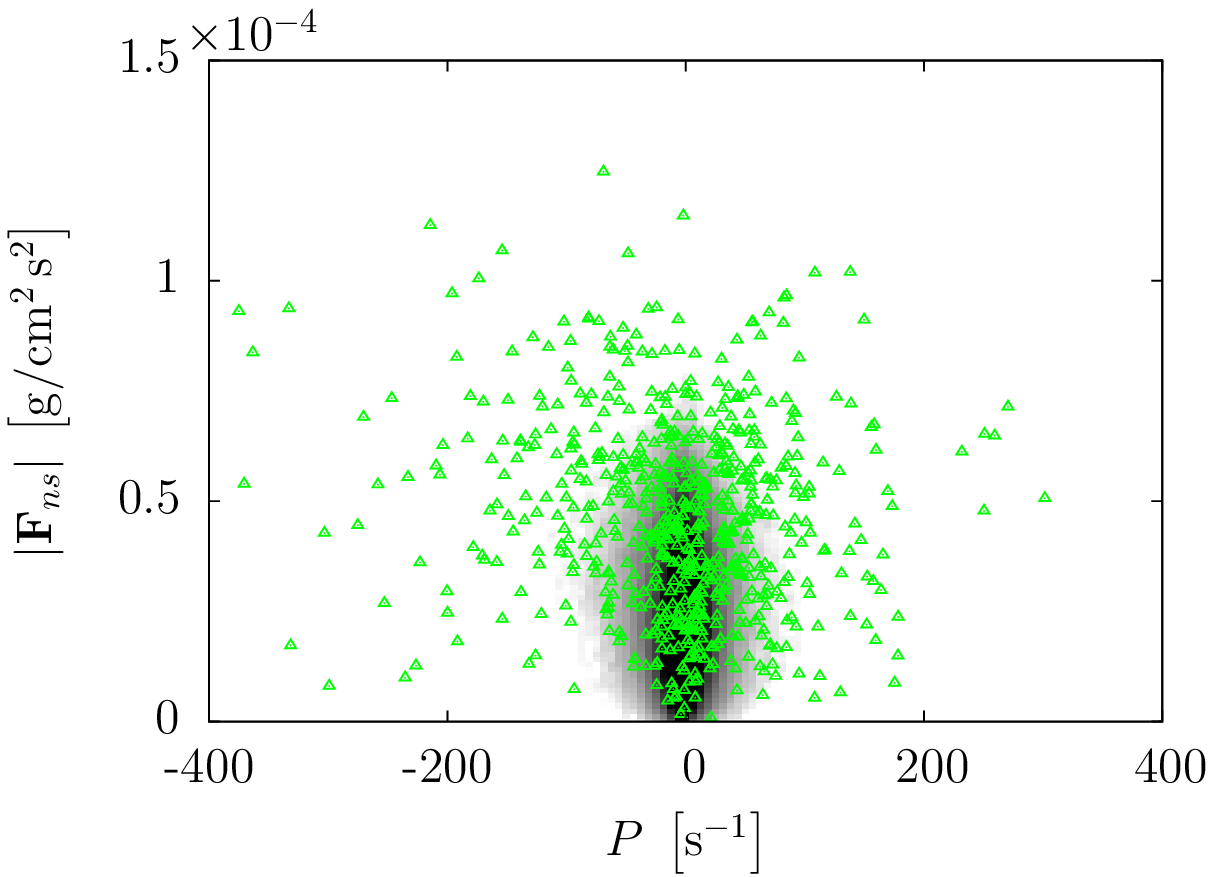}
	
	\caption{Kernel density estimates of the probability density function of the mutual friction ${\bf F}_{ns}$ verses the vorticity production $P=-\kappa \dot{\bs}\cdot \bn$ for the Vinen tangle (left), Counterflow tangle (centre) and Polarized tangle (right). The coloured symbols indicate reconnection events.}
	\label{fig:pdf_mf_vort}
\end{figure}

The last set of PDFs shown in Fig.~\ref{fig:pdf_mf_vort} display the relationship between the mutual friction and the vorticity production rate $P$ defined as
\begin{equation}
P(\xi) = - \kappa \dot{\bs}\cdot \bn, 
\end{equation}
where $\bn$ is the local normal vector to the vortex line. $P$ quantifies the local rate in which the vortex line length increases or decreases, and therefore can be used as a measure of vorticity production in quantum turbulence.  We observe in Fig.~\ref{fig:pdf_mf_vort} that the majority of the points lie close to the zero vorticity production rate region, indicating that many areas of the tangle are neither stretching nor contracting. However, what we do find is that many reconnection events appear to reduce the vortex line length corresponding to negative $P$.  This feature is most pronounced in the Vinen and Counterflow tangles. The most effective way for a vortex ring to increase in size is to propagate in the direction of the normal fluid flow. In steady state conditions this process in balanced by the dissipative effects of mutual friction. However, reconnections can lead to topological changes in the local structure of the vortex line and randomize the direction of the vortex segment. In turn, this may greatly decrease its ability to absorb energy from the fluid flow, consequently, leading to the mutual friction dominating and the vortex line to shrink. One may also expect that this process will be enhanced if the reconnections are of the larger angle variety as they will produce larger curvatures and larger velocities.  This seems to be the case indicated by the values of $P$ for the reconnection events given in Table~\ref{tab:mean}.

\section{Conclusions and discussions}
\label{sec:conclusions}

We have performed numerical simulations using the vortex filament model for finite temperature quantum turbulence, and have considered three physically relevant types of vortex tangles in order to investigate how the tangle structure and vortex reconnection dynamics effect the behaviour of the mutual friction force.

We have found that quantized vortex reconnections lead to areas of high curvature and subsequently to regions of intense superfluid velocity.  Hence in isotropic tangles, where the mean mutual friction force is close to zero, vortex reconnections are primarily responsible for areas of high mutual friction. We also note that high polarization of the tangle can lead to a suppression of vortex reconnections reducing the frequency of high mutual friction `explosions'. 

In contrast, for superfluid counterflow, we observe that the imposed normal fluid leads to a significant mean mutual friction force.  Vortex reconnections lead to further fluctuations on the background force, and so can reduce the local value of the mutual friction if they lead to velocity fluctuations aligned with the normal fluid flow.  However, it is worth remarking that the most extreme values of mutual friction are associated with vortex reconnections.
If one considers a more realistic setup with solid boundaries, it has been shown~\cite{baggaley_thermal_2014} that the vortex line density is not homogeneous. As the reconnection rate is proportional to the vortex line density~\cite{poole_geometry_2003}, one can expect that in superfluid counterflow, the specific spatial structure of the vortex tangle will play an essential role in the characteristics of the mutual friction force.

Furthermore, it is important to appreciate that computational limitations (resolution) severely constrain the maximum curvature that one can resolve in numerical simulations performed with the vortex filament mode. From Figs.~\ref{fig:pdf_mf_curv} and \ref{fig:pdf_curv_vel} we can visibly see that curvatures are capped by our resolution at $\kappa_{\rm max} \simeq 1250~{\rm cm}^{-1}$; in reality we would expect quantized vortex reconnections to create curvatures many orders of magnitude larger than the results of the numerics presented here, and hence produce larger values of mutual friction force.  This will only enhance the importance quantized vortex reconnections has upon the mutual friction force.

As is indicated by our numerical simulations, if vortex reconnections are indeed important events that lead to the largest values of the mutual friction force, then one should try to understand how such small-scale, localized regions, will effect the normal fluid flow.  Indeed, as our analysis indicates, the mutual friction forcing due to reconnections will effect the normal fluid at extremely small-scales; possibly scales far smaller than the viscous dissipation scale. One could imagine that adding a forcing term acting at very small scales to the equation for the normal fluid component in the coarse-grained HVBK equations (see~\cite{roche_quantum_2009}) to approximate vortex reconnections would be appropriate, although at such small scales numerical instabilities will pose a problem. Furthermore, we would not expect this to effect the large-scale dynamics of the turbulence in the normal fluid, as the forcing would occur on scales many orders of magnitudes smaller than the integral scale. However, Lagrangian quantities may well be affected by the mutual friction force on the normal fluid. In particular recent experimental studies~\cite{paoletti_velocity_2008,la_mantia_lagrangian_2013} have made use of particle tracking techniques to probe Lagrangian quantities such as one-point velocity and acceleration statistics. For particles following the normal fluid one may expect that the one-point velocity statistics to follow a Gaussian distribution, as observed in classical turbulence~\cite{vincent_spatial_1991}. Our results here suggest that this may not be the case as large fluctuations in the particles velocity and acceleration could occur due to the effect of quantized vortex reconnections on the normal fluid.

\acknowledgements
We would like to thank Sergey Nazarenko for the initial discussions associated to this work, and thank Risto H\"{a}nninen for fruitful communications.

\bibliographystyle{apsrev}
\bibliography{biblio}

\begin{thebibliography}{33}
\expandafter\ifx\csname natexlab\endcsname\relax\def\natexlab#1{#1}\fi
\expandafter\ifx\csname bibnamefont\endcsname\relax
  \def\bibnamefont#1{#1}\fi
\expandafter\ifx\csname bibfnamefont\endcsname\relax
  \def\bibfnamefont#1{#1}\fi
\expandafter\ifx\csname citenamefont\endcsname\relax
  \def\citenamefont#1{#1}\fi
\expandafter\ifx\csname url\endcsname\relax
  \def\url#1{\texttt{#1}}\fi
\expandafter\ifx\csname urlprefix\endcsname\relax\def\urlprefix{URL }\fi
\providecommand{\bibinfo}[2]{#2}
\providecommand{\eprint}[2][]{\url{#2}}

\bibitem[{\citenamefont{Tisza}(1947)}]{tisza_theory_1947}
\bibinfo{author}{\bibfnamefont{L.}~\bibnamefont{Tisza}},
  \bibinfo{journal}{Phys. Rev.} \textbf{\bibinfo{volume}{72}},
  \bibinfo{pages}{838} (\bibinfo{year}{1947}).

\bibitem[{\citenamefont{Landau}(1941)}]{landau_theory_1941}
\bibinfo{author}{\bibfnamefont{L.}~\bibnamefont{Landau}},
  \bibinfo{journal}{Phys. Rev.} \textbf{\bibinfo{volume}{60}},
  \bibinfo{pages}{356} (\bibinfo{year}{1941}).

\bibitem[{\citenamefont{Landau}(1949)}]{landau_theory_1949}
\bibinfo{author}{\bibfnamefont{L.}~\bibnamefont{Landau}},
  \bibinfo{journal}{Phys. Rev.} \textbf{\bibinfo{volume}{75}},
  \bibinfo{pages}{884} (\bibinfo{year}{1949}).

\bibitem[{\citenamefont{Vinen and Niemela}(2002)}]{vinen_quantum_2002}
\bibinfo{author}{\bibfnamefont{W.~F.} \bibnamefont{Vinen}} \bibnamefont{and}
  \bibinfo{author}{\bibfnamefont{J.~J.} \bibnamefont{Niemela}},
  \bibinfo{journal}{J. Low Temp. Phys.} \textbf{\bibinfo{volume}{128}},
  \bibinfo{pages}{167} (\bibinfo{year}{2002}), ISSN \bibinfo{issn}{0022-2291,
  1573-7357}.

\bibitem[{\citenamefont{Vinen}(2006)}]{vinen_introduction_2006}
\bibinfo{author}{\bibfnamefont{W.~F.} \bibnamefont{Vinen}},
  \bibinfo{journal}{J. Low Temp. Phys.} \textbf{\bibinfo{volume}{145}},
  \bibinfo{pages}{7} (\bibinfo{year}{2006}), ISSN \bibinfo{issn}{0022-2291,
  1573-7357}.

\bibitem[{\citenamefont{Frisch}(1995)}]{frisch_turbulence:_1995}
\bibinfo{author}{\bibfnamefont{U.}~\bibnamefont{Frisch}},
  \emph{\bibinfo{title}{Turbulence: The Legacy of A. N. Kolmogorov}}
  (\bibinfo{publisher}{Cambridge University Press}, \bibinfo{year}{1995}), ISBN
  \bibinfo{isbn}{9780521457132}.

\bibitem[{\citenamefont{Baggaley}(2012)}]{baggaley_importance_2012}
\bibinfo{author}{\bibfnamefont{A.~W.} \bibnamefont{Baggaley}},
  \bibinfo{journal}{Phys. Fluids} \textbf{\bibinfo{volume}{24}},
  \bibinfo{pages}{055109} (\bibinfo{year}{2012}), ISSN
  \bibinfo{issn}{1070-6631, 1089-7666}.

\bibitem[{\citenamefont{Baggaley et~al.}(2012)\citenamefont{Baggaley, Laurie,
  and Barenghi}}]{baggaley_vortex-density_2012}
\bibinfo{author}{\bibfnamefont{A.~W.} \bibnamefont{Baggaley}},
  \bibinfo{author}{\bibfnamefont{J.}~\bibnamefont{Laurie}}, \bibnamefont{and}
  \bibinfo{author}{\bibfnamefont{C.~F.} \bibnamefont{Barenghi}},
  \bibinfo{journal}{Phys. Rev. Lett.} \textbf{\bibinfo{volume}{109}},
  \bibinfo{pages}{205304} (\bibinfo{year}{2012}).

\bibitem[{\citenamefont{Adachi et~al.}(2010)\citenamefont{Adachi, Fujiyama, and
  Tsubota}}]{adachi_steady-state_2010}
\bibinfo{author}{\bibfnamefont{H.}~\bibnamefont{Adachi}},
  \bibinfo{author}{\bibfnamefont{S.}~\bibnamefont{Fujiyama}}, \bibnamefont{and}
  \bibinfo{author}{\bibfnamefont{M.}~\bibnamefont{Tsubota}},
  \bibinfo{journal}{Phys. Rev. B} \textbf{\bibinfo{volume}{81}},
  \bibinfo{pages}{104511} (\bibinfo{year}{2010}).

\bibitem[{\citenamefont{Hall and Vinen}(1956)}]{hall_rotation_1956}
\bibinfo{author}{\bibfnamefont{H.~E.} \bibnamefont{Hall}} \bibnamefont{and}
  \bibinfo{author}{\bibfnamefont{W.~F.} \bibnamefont{Vinen}},
  \bibinfo{journal}{Proc. R. Soc. Lond. A} \textbf{\bibinfo{volume}{238}},
  \bibinfo{pages}{215} (\bibinfo{year}{1956}), ISSN \bibinfo{issn}{1364-5021,
  1471-2946}.

\bibitem[{\citenamefont{Bekarevich and
  Khalatnikov}(1961)}]{bekarevich_phenomenological_1961}
\bibinfo{author}{\bibfnamefont{I.}~\bibnamefont{Bekarevich}} \bibnamefont{and}
  \bibinfo{author}{\bibfnamefont{I.}~\bibnamefont{Khalatnikov}},
  \bibinfo{journal}{J. Exp. Theor. Phys.} \textbf{\bibinfo{volume}{13}},
  \bibinfo{pages}{643} (\bibinfo{year}{1961}).

\bibitem[{\citenamefont{Henderson et~al.}(1995)\citenamefont{Henderson,
  Barenghi, and Jones}}]{henderson_nonlinear_1995}
\bibinfo{author}{\bibfnamefont{K.~L.} \bibnamefont{Henderson}},
  \bibinfo{author}{\bibfnamefont{C.~F.} \bibnamefont{Barenghi}},
  \bibnamefont{and} \bibinfo{author}{\bibfnamefont{C.~A.} \bibnamefont{Jones}},
  \bibinfo{journal}{J. Fluid Mech.} \textbf{\bibinfo{volume}{283}},
  \bibinfo{pages}{329} (\bibinfo{year}{1995}), ISSN \bibinfo{issn}{1469-7645}.

\bibitem[{\citenamefont{Roche et~al.}(2009)\citenamefont{Roche, Barenghi, and
  Leveque}}]{roche_quantum_2009}
\bibinfo{author}{\bibfnamefont{P.-E.} \bibnamefont{Roche}},
  \bibinfo{author}{\bibfnamefont{C.~F.} \bibnamefont{Barenghi}},
  \bibnamefont{and} \bibinfo{author}{\bibfnamefont{E.}~\bibnamefont{Leveque}},
  \bibinfo{journal}{{EPL}} \textbf{\bibinfo{volume}{87}},
  \bibinfo{pages}{54006} (\bibinfo{year}{2009}), ISSN
  \bibinfo{issn}{0295-5075}.

\bibitem[{\citenamefont{Bewley et~al.}(2008)\citenamefont{Bewley, Paoletti,
  Sreenivasan, and Lathrop}}]{bewley_characterization_2008}
\bibinfo{author}{\bibfnamefont{G.~P.} \bibnamefont{Bewley}},
  \bibinfo{author}{\bibfnamefont{M.~S.} \bibnamefont{Paoletti}},
  \bibinfo{author}{\bibfnamefont{K.~R.} \bibnamefont{Sreenivasan}},
  \bibnamefont{and} \bibinfo{author}{\bibfnamefont{D.~P.}
  \bibnamefont{Lathrop}}, \bibinfo{journal}{{PNAS}}
  \textbf{\bibinfo{volume}{105}}, \bibinfo{pages}{13707}
  (\bibinfo{year}{2008}), ISSN \bibinfo{issn}{0027-8424, 1091-6490}.

\bibitem[{\citenamefont{Skrbek and Sreenivasan}(2012)}]{skrbek_developed_2012}
\bibinfo{author}{\bibfnamefont{L.}~\bibnamefont{Skrbek}} \bibnamefont{and}
  \bibinfo{author}{\bibfnamefont{K.~R.} \bibnamefont{Sreenivasan}},
  \bibinfo{journal}{Phys. Fluids} \textbf{\bibinfo{volume}{24}},
  \bibinfo{pages}{011301} (\bibinfo{year}{2012}), ISSN
  \bibinfo{issn}{1070-6631, 1089-7666}.

\bibitem[{\citenamefont{Kivotides et~al.}(2000)\citenamefont{Kivotides,
  Barenghi, and Samuels}}]{kivotides_triple_2000}
\bibinfo{author}{\bibfnamefont{D.}~\bibnamefont{Kivotides}},
  \bibinfo{author}{\bibfnamefont{C.~F.} \bibnamefont{Barenghi}},
  \bibnamefont{and} \bibinfo{author}{\bibfnamefont{D.~C.}
  \bibnamefont{Samuels}}, \bibinfo{journal}{Science}
  \textbf{\bibinfo{volume}{290}}, \bibinfo{pages}{777} (\bibinfo{year}{2000}),
  ISSN \bibinfo{issn}{0036-8075, 1095-9203}.

\bibitem[{\citenamefont{Kivotides et~al.}(2001)\citenamefont{Kivotides,
  Barenghi, and Samuels}}]{kivotides_superfluid_2001}
\bibinfo{author}{\bibfnamefont{D.}~\bibnamefont{Kivotides}},
  \bibinfo{author}{\bibfnamefont{C.~F.} \bibnamefont{Barenghi}},
  \bibnamefont{and} \bibinfo{author}{\bibfnamefont{D.~C.}
  \bibnamefont{Samuels}}, \bibinfo{journal}{{EPL}}
  \textbf{\bibinfo{volume}{54}}, \bibinfo{pages}{774} (\bibinfo{year}{2001}),
  ISSN \bibinfo{issn}{0295-5075}.

\bibitem[{\citenamefont{H{\"a}nninen}(2013)}]{hanninen_dissipation_2013}
\bibinfo{author}{\bibfnamefont{R.}~\bibnamefont{H{\"a}nninen}},
  \bibinfo{journal}{Phys. Rev. B} \textbf{\bibinfo{volume}{88}},
  \bibinfo{pages}{054511} (\bibinfo{year}{2013}).

\bibitem[{\citenamefont{Schwarz}(1985)}]{schwarz_three-dimensional_1985}
\bibinfo{author}{\bibfnamefont{K.~W.} \bibnamefont{Schwarz}},
  \bibinfo{journal}{Phys. Rev. B} \textbf{\bibinfo{volume}{31}},
  \bibinfo{pages}{5782} (\bibinfo{year}{1985}).

\bibitem[{\citenamefont{Saffman}(1992)}]{saffman_vortex_1992}
\bibinfo{author}{\bibfnamefont{P.~G.} \bibnamefont{Saffman}},
  \emph{\bibinfo{title}{Vortex Dynamics}} (\bibinfo{publisher}{Cambridge
  University Press}, \bibinfo{year}{1992}), ISBN \bibinfo{isbn}{9780521420587}.

\bibitem[{\citenamefont{Donnelly and Barenghi}(1998)}]{donnelly_observed_1998}
\bibinfo{author}{\bibfnamefont{R.~J.} \bibnamefont{Donnelly}} \bibnamefont{and}
  \bibinfo{author}{\bibfnamefont{C.~F.} \bibnamefont{Barenghi}},
  \bibinfo{journal}{J. Phys. Chem. Ref. Data} \textbf{\bibinfo{volume}{27}},
  \bibinfo{pages}{1217} (\bibinfo{year}{1998}), ISSN \bibinfo{issn}{0047-2689,
  1529-7845}.

\bibitem[{\citenamefont{Idowu et~al.}(2000)\citenamefont{Idowu, Kivotides,
  Barenghi, and Samuels}}]{idowu_equation_2000}
\bibinfo{author}{\bibfnamefont{O.~C.} \bibnamefont{Idowu}},
  \bibinfo{author}{\bibfnamefont{D.}~\bibnamefont{Kivotides}},
  \bibinfo{author}{\bibfnamefont{C.~F.} \bibnamefont{Barenghi}},
  \bibnamefont{and} \bibinfo{author}{\bibfnamefont{D.~C.}
  \bibnamefont{Samuels}}, \bibinfo{journal}{J. Low Temp. Phys.}
  \textbf{\bibinfo{volume}{120}}, \bibinfo{pages}{269} (\bibinfo{year}{2000}),
  ISSN \bibinfo{issn}{0022-2291, 1573-7357}.

\bibitem[{\citenamefont{Baggaley and
  Barenghi}(2011)}]{baggaley_vortex-density_2011}
\bibinfo{author}{\bibfnamefont{A.~W.} \bibnamefont{Baggaley}} \bibnamefont{and}
  \bibinfo{author}{\bibfnamefont{C.~F.} \bibnamefont{Barenghi}},
  \bibinfo{journal}{Phys. Rev. B} \textbf{\bibinfo{volume}{84}},
  \bibinfo{pages}{020504} (\bibinfo{year}{2011}).

\bibitem[{\citenamefont{Baggaley and Barenghi}(2012)}]{baggaley_tree_2012}
\bibinfo{author}{\bibfnamefont{A.~W.} \bibnamefont{Baggaley}} \bibnamefont{and}
  \bibinfo{author}{\bibfnamefont{C.~F.} \bibnamefont{Barenghi}},
  \bibinfo{journal}{J. Low Temp. Phys.} \textbf{\bibinfo{volume}{166}},
  \bibinfo{pages}{3} (\bibinfo{year}{2012}), ISSN \bibinfo{issn}{0022-2291,
  1573-7357}.

\bibitem[{\citenamefont{Li et~al.}(2008)\citenamefont{Li, Perlman, Wan, Yang,
  Meneveau, Burns, Chen, Szalay, and Eyink}}]{li_public_2008}
\bibinfo{author}{\bibfnamefont{Y.}~\bibnamefont{Li}},
  \bibinfo{author}{\bibfnamefont{E.}~\bibnamefont{Perlman}},
  \bibinfo{author}{\bibfnamefont{M.}~\bibnamefont{Wan}},
  \bibinfo{author}{\bibfnamefont{Y.}~\bibnamefont{Yang}},
  \bibinfo{author}{\bibfnamefont{C.}~\bibnamefont{Meneveau}},
  \bibinfo{author}{\bibfnamefont{R.}~\bibnamefont{Burns}},
  \bibinfo{author}{\bibfnamefont{S.}~\bibnamefont{Chen}},
  \bibinfo{author}{\bibfnamefont{A.}~\bibnamefont{Szalay}}, \bibnamefont{and}
  \bibinfo{author}{\bibfnamefont{G.}~\bibnamefont{Eyink}}, \bibinfo{journal}{J.
  Turbulence} p. \bibinfo{pages}{N31} (\bibinfo{year}{2008}).

\bibitem[{\citenamefont{Kivotides}(2006)}]{kivotides_coherent_2006}
\bibinfo{author}{\bibfnamefont{D.}~\bibnamefont{Kivotides}},
  \bibinfo{journal}{Phys. Rev. Lett.} \textbf{\bibinfo{volume}{96}},
  \bibinfo{pages}{175301} (\bibinfo{year}{2006}).

\bibitem[{\citenamefont{L{\textquoteright}vov
  et~al.}(2007)\citenamefont{L{\textquoteright}vov, Nazarenko, and
  Rudenko}}]{lvov_bottleneck_2007}
\bibinfo{author}{\bibfnamefont{V.~S.} \bibnamefont{L{\textquoteright}vov}},
  \bibinfo{author}{\bibfnamefont{S.~V.} \bibnamefont{Nazarenko}},
  \bibnamefont{and} \bibinfo{author}{\bibfnamefont{O.}~\bibnamefont{Rudenko}},
  \bibinfo{journal}{Phys. Rev. B} \textbf{\bibinfo{volume}{76}},
  \bibinfo{pages}{024520} (\bibinfo{year}{2007}).

\bibitem[{\citenamefont{Silverman}(1986)}]{silverman_density_1986}
\bibinfo{author}{\bibfnamefont{B.~W.} \bibnamefont{Silverman}},
  \emph{\bibinfo{title}{Density Estimation for Statistics and Data Analysis}}
  (\bibinfo{publisher}{{CRC} Press}, \bibinfo{year}{1986}), ISBN
  \bibinfo{isbn}{9780412246203}.

\bibitem[{\citenamefont{Baggaley and Laurie}(2014)}]{baggaley_thermal_2014}
\bibinfo{author}{\bibfnamefont{A.~W.} \bibnamefont{Baggaley}} \bibnamefont{and}
  \bibinfo{author}{\bibfnamefont{J.}~\bibnamefont{Laurie}},
  \bibinfo{journal}{J. Low. Temp. Phys.} pp. \bibinfo{pages}{1--18}
  (\bibinfo{year}{2014}), ISSN \bibinfo{issn}{0022-2291, 1573-7357}.

\bibitem[{\citenamefont{Poole et~al.}(2003)\citenamefont{Poole, Scoffield,
  Barenghi, and Samuels}}]{poole_geometry_2003}
\bibinfo{author}{\bibfnamefont{D.~R.} \bibnamefont{Poole}},
  \bibinfo{author}{\bibfnamefont{H.}~\bibnamefont{Scoffield}},
  \bibinfo{author}{\bibfnamefont{C.~F.} \bibnamefont{Barenghi}},
  \bibnamefont{and} \bibinfo{author}{\bibfnamefont{D.~C.}
  \bibnamefont{Samuels}}, \bibinfo{journal}{J. Low Temp. Phys.}
  \textbf{\bibinfo{volume}{132}}, \bibinfo{pages}{97} (\bibinfo{year}{2003}),
  ISSN \bibinfo{issn}{0022-2291, 1573-7357}.

\bibitem[{\citenamefont{Paoletti et~al.}(2008)\citenamefont{Paoletti, Fisher,
  Sreenivasan, and Lathrop}}]{paoletti_velocity_2008}
\bibinfo{author}{\bibfnamefont{M.~S.} \bibnamefont{Paoletti}},
  \bibinfo{author}{\bibfnamefont{M.~E.} \bibnamefont{Fisher}},
  \bibinfo{author}{\bibfnamefont{K.~R.} \bibnamefont{Sreenivasan}},
  \bibnamefont{and} \bibinfo{author}{\bibfnamefont{D.~P.}
  \bibnamefont{Lathrop}}, \bibinfo{journal}{Phys. Rev. Lett.}
  \textbf{\bibinfo{volume}{101}}, \bibinfo{pages}{154501}
  (\bibinfo{year}{2008}).

\bibitem[{\citenamefont{La~Mantia et~al.}(2013)\citenamefont{La~Mantia, Duda,
  Rotter, and Skrbek}}]{la_mantia_lagrangian_2013}
\bibinfo{author}{\bibfnamefont{M.}~\bibnamefont{La~Mantia}},
  \bibinfo{author}{\bibfnamefont{D.}~\bibnamefont{Duda}},
  \bibinfo{author}{\bibfnamefont{M.}~\bibnamefont{Rotter}}, \bibnamefont{and}
  \bibinfo{author}{\bibfnamefont{L.}~\bibnamefont{Skrbek}},
  \bibinfo{journal}{J. Fluid Mech.} \textbf{\bibinfo{volume}{717}}
  (\bibinfo{year}{2013}), ISSN \bibinfo{issn}{1469-7645}.

\bibitem[{\citenamefont{Vincent and Meneguzzi}(1991)}]{vincent_spatial_1991}
\bibinfo{author}{\bibfnamefont{A.}~\bibnamefont{Vincent}} \bibnamefont{and}
  \bibinfo{author}{\bibfnamefont{M.}~\bibnamefont{Meneguzzi}},
  \bibinfo{journal}{J. Fluid Mech.} \textbf{\bibinfo{volume}{225}},
  \bibinfo{pages}{1} (\bibinfo{year}{1991}), ISSN \bibinfo{issn}{1469-7645}.

\end{thebibliography}

\end{document}